\documentclass[%
 reprint,
superscriptaddress,
 amsmath,amssymb,
 aps,
]{revtex4-2}

\usepackage{amsmath}
\usepackage{graphicx}
\usepackage{dcolumn}
\usepackage{bm}
\usepackage{xcolor}
\usepackage{mhchem}
\usepackage{xr}
\usepackage{filecontents}
\usepackage{multirow}
\usepackage{array}
\usepackage{comment}

\makeatletter

\newcommand*{\addFileDependency}[1]{
\typeout{(#1)}

\@addtofilelist{#1}

\IfFileExists{#1}{}{\typeout{No file #1.}}
}\makeatother

\newcommand*{\myexternaldocument}[1]{%
\externaldocument{#1}%
\addFileDependency{#1.tex}%
\addFileDependency{#1.aux}%
}

\myexternaldocument{SI}

\makeatletter
\newcommand*{\balancecolsandclearpage}{%
  \close@column@grid
  \clearpage
  \twocolumngrid
}
\makeatother

\newcommand{\bp}{\mathbf{p}}
\newcommand{\cW}{\mathcal{W}}
\newcommand{\cR}{\mathcal{R}}

\begin{document}

\preprint{APS/123-QED}

\title{A topological mechanism for robust and efficient global oscillations in biological networks}

\author{Chongbin Zheng}
\affiliation{
 Department of Physics and Astronomy, Rice University, Houston, Texas 77005, USA
}
\affiliation{
 Center for Theoretical Biological Physics, Rice University, Houston, Texas 77005, USA
}
\author{Evelyn Tang}
\affiliation{
 Department of Physics and Astronomy, Rice University, Houston, Texas 77005, USA
}
\affiliation{
 Center for Theoretical Biological Physics, Rice University, Houston, Texas 77005, USA
}

\begin{abstract}
Long and stable timescales are often observed in complex biochemical networks, such as in emergent oscillations. How these robust dynamics persist remains unclear, given the many stochastic reactions and shorter time scales demonstrated by underlying components. We propose a topological model that produces long oscillations around the network boundary, reducing the system dynamics to a lower-dimensional current in a robust manner. Using this to model KaiC, which regulates the circadian rhythm in cyanobacteria, we compare the coherence of oscillations to that in other KaiC models. Our topological model localizes currents on the system edge, with an efficient regime of simultaneously increased precision and decreased cost. Further, we introduce a new predictor of coherence from the analysis of spectral gaps, and show that our model saturates a global thermodynamic bound. Our work presents a new mechanism and parsimonious description 
for robust emergent oscillations in complex biological networks.
\end{abstract}

\maketitle

\section{Introduction}
The reduction of the full system response to a much lower dimensional description has been observed in many complex biological systems, where the system dynamics or behavior reduce to a much smaller phase space \cite{rigotti2013importance, tang2019effective, stephens2008dimensionality}. However, we still lack good models that can mechanistically account for this dimensionality reduction, or that remain stable under noise or structural heterogeneity. This is exemplified in computational models of memory, that describe specific attractor states which represent persistent memories \cite{chaudhuri2016computational, hopfield1982neural}. However, attractors tend to drift or lose accuracy with noise and it remains an area of open research on how to retain encoded information in these models \cite{chaudhuri2016computational}. Another example is that of long oscillations, such as the circadian rhythm, which are crucial for the regulation of many processes such as metabolism and replication \cite{liao2021circadian, puszynska2017switching}. Previously proposed models such as feedback loops of chemical reactions typically involve either small reaction networks consisting of reactions on a similar timescale as the oscillation itself, or a large number of system-dependent parameters \cite{smolen2001modeling,van2007allosteric,paijmans2017thermodynamically}. Simple conditions are still lacking for explaining how such oscillations with their long timescales can emerge from a large phase space of faster chemical reactions. Understanding the necessary and sufficient principles that govern robust oscillations is crucial as disruptions in biological clocks lead to decreased health and reproductive fitness in multiple organisms \cite{puszynska2017switching, green2002circadian, horn2019circadian,scheer2009adverse, masri2018emerging}.

As biological networks including those mentioned above typically have a large phase space of possible reactions, this renders unfeasible exhaustive searches using other approaches like experiments or numerical simulation \cite{dewyer2018methods}, underscoring the need for simple conceptual methods to provide insight \cite{davies2020does, winfree1980geometry, gao2015simplicity}. The development of rigorous theory would also shed light on simple design principles for targeted dynamics in synthetic biological systems \cite{doudna2017will, schwille2018maxsynbio} or in the engineering of reconfigurable materials, e.g., through dissipative self-assembly \cite{deng2020atp, riess2020design}. Yet, biology presents challenges for the development of suitable theory due to being stochastic, heterogeneous, and strongly non-equilibrium \cite{ross2009complex, ashkenasy2017systems, dewyer2018methods, qian2021stochastic}. Hence, the few successful models that exist in biology are often heavily dependent on specific system parameters.  

\begin{figure*}[t]
	\centering
	\includegraphics[width=15cm, height=11.3cm]{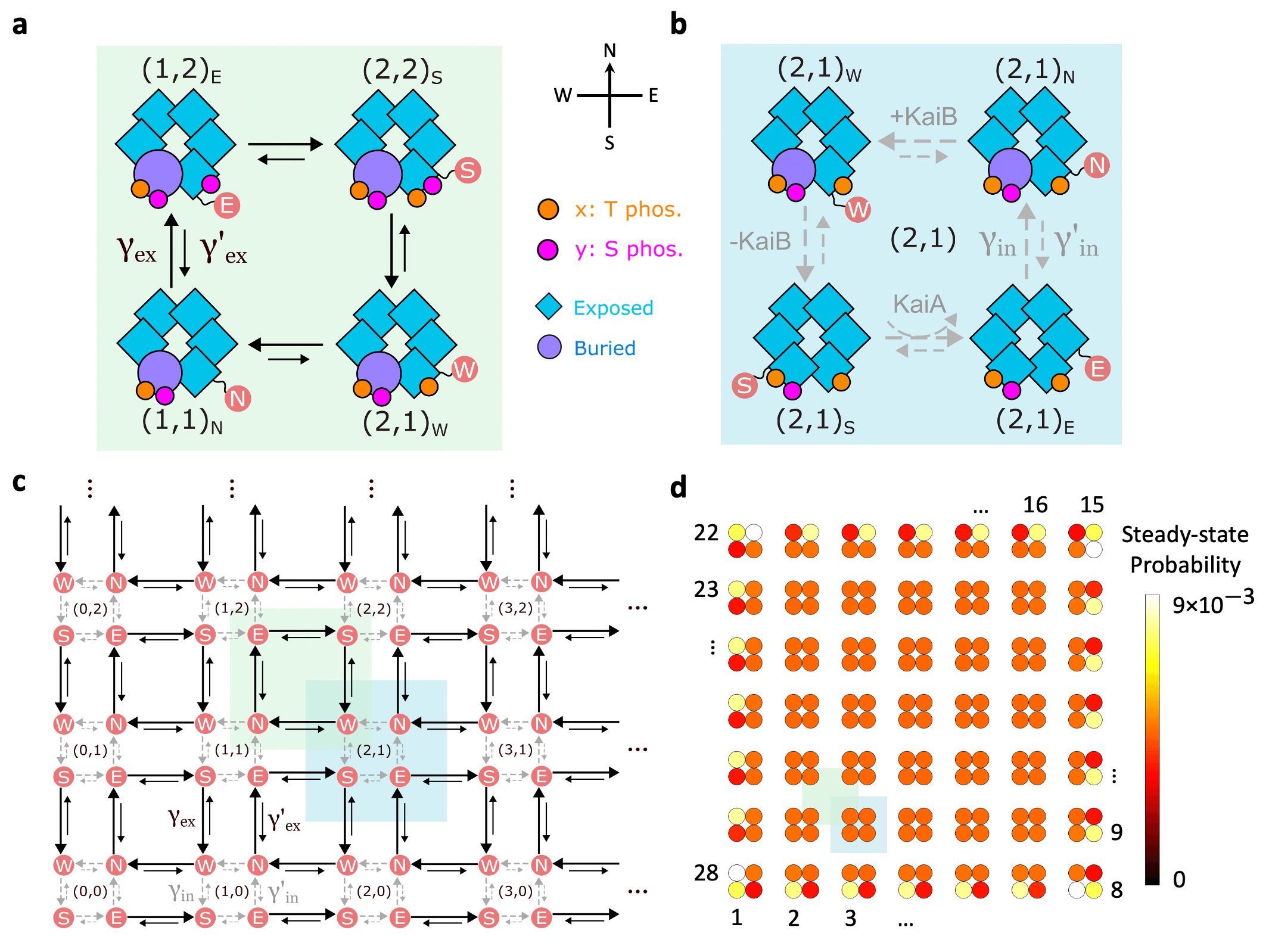}
	\caption{\textbf{Topological model for emergent oscillations, illustrated with KaiABC that regulates the circadian rhythm.} \textbf{a}, Based on observations of autophosphorylation in the literature \cite{brettschneider2010sequestration,kageyama2006cyanobacterial,lin2014mixtures}, it is thought that monomers undergo phosphorylation and dephosphorylation cycles (black arrows $\gamma_{\text{ex}}$ and slower reverse transitions $\gamma_{\text{ex}}'$). Two types of phosphorylation T and S are shown with the addition of orange and pink circles and the numbers of each are given in brackets. \textbf{b}, Within a given phosphorylation level, internal transitions (gray arrows, $\gamma_{\text{in}}$ and $\gamma_{\text{in}}'$)  take place due to conformational changes (illustrated by circles and squares) or interaction with proteins like KaiA or KaiB. The internal state labels the direction of the next external transition that it primes. KaiA promotes phosphorylation catalytically \cite{kageyama2006cyanobacterial} and hence is denoted with a curved arrow. \textbf{c}, These cycles can be laid out in a lattice, with T phosphorylation along the horizontal axis and S phosphorylation along the vertical axis, where each set of four internal transitions (highlighted in blue) repeats along these axes. \textbf{d}, This lattice allows probing of its topological properties. In the ordinary case with similar rates throughout, i.e. $\gamma_{\text{ex}}\sim\gamma_{\text{in}}$, the system will perform a random walk ergodically through the phase space. In the topological regime when $\gamma_{\text{ex}}\gg\gamma_{\text{in}}$, once the system hits an edge it will continue around the edge, as can be verified by inspection. We plot the steady state in the topological regime which lies on the system edge, taking $\mu=1,\rho=2,\gamma_{\text{tot}}=1$. There are 28 states along the edge with high probability, labeled by the order in which they are traversed in a typical trajectory. These trajectories form a global current along the edge of the state space.}
	\label{plt_model}
\end{figure*} 

Towards dimensional reduction onto the network boundary, topological models have been proposed in stochastic systems \cite{murugan2017topologically,dasbiswas2018topological,tang2021topology}. These are a generalization of topological invariants studied in quantum systems \cite{moore2010birth, chiu2016classification, stormer1999fractional, tang2012interacting}, which show physical responses on the system edge or boundary. Powerfully, this response is insensitive to various types of disorder or noise \cite{weis2011metrology, murugan2017topologically,tang2021topology}.  It would be desirable to demonstrate how topology can be realized in a biological system, given the many attractive properties of topology such as its robust response.  However, it has not yet been shown to relate to a biological system. Previous works do not explicitly connect to experimentally tunable parameters or known molecular reactions, giving little guidance on how to measure or access the topological properties \cite{murugan2017topologically,dasbiswas2018topological,tang2021topology}. Our manuscript responds to this lack in the field, by providing detailed biophysical mechanisms towards experimental verification of topological invariants.

Here, we provide the first case study of a topological mechanism in a concrete biological system -- that of the KaiABC system for the circadian rhythm of cyanobacteria \cite{ishiura1998expression,nakajima2005reconstitution}. The topological edge currents naturally reproduce the kinetic ordering of KaiC phosphorylation cycles, first proposed in \cite{tang2021topology} but with the biophysical mechanisms remaining abstract. These missing mechanisms are made concrete in this work, where we demonstrate how the biochemical reactions \cite{kim2008day,chang2011flexibility,tseng2014cooperative,hong2018molecular} interact via a separation of timescales to produce the resulting edge currents. This provides the necessary and sufficient conditions for the observed oscillatory cycle. We analyze the parameter space to show how a variation of the transition rates (e.g. by changing ATP concentration \cite{phong2013robust} or using a mutant \cite{qin2010intermolecular}) affects the coherence and dissipation of the oscillation, by tuning the system into and out of the topological transition. This yields key insights into an important regulatory system that had until now required rather complicated models, especially to reproduce the observed kinetic ordering of T and S phosphorylation \cite{van2007allosteric,paijmans2017thermodynamically,smolen2001modeling}. 

On the theoretical level, we characterize the coherence of the resulting cycle, showing that it satisfies theoretical bounds \cite{barato2017coherence} for the most coherent oscillator equivalent to that of a unicycle network – without the fine-tuning needed for a unicycle \cite{barato2017coherence}. The topological model shows high coherence compared to other available models, while producing the global day-night cycle with unusually few free parameters. In addition, we explore the coherence and energetic cost of the oscillation using tools from non-equilibrium stochastic thermodynamics, to reveal an efficient regime where coherence increases while cost simultaneously decreases. Lastly, a new indicator of oscillation coherence from spectral gaps in band theory is introduced, to study the saturation of this model on global thermodynamic bounds. Overall, this analysis can explain long-standing puzzles in biology such as how dimensional reduction is achieved in a robust and flexible manner to produce emergent robust oscillations.

Our work provides an alternative mechanism for oscillations from the current paradigm from Monod-Wyman-Changeux (MWC) \cite{monod1965nature}. This paradigm assumes cooperative all-or-none conformational changes for protomers in an oligomer upon ligand binding, which acts as a molecular switch that changes the affinity of all binding sites. The model has been useful in describing systems such as hemoglobin \cite{monod1965nature}, ligand-gated ion channels \cite{calimet2013gating}, and bacterial chemotaxis \cite{mello2005allosteric}. Still, it remains unclear if the MWC model is the dominant mechanism for other systems. Indeed, models for KaiABC typically assume highly cooperative conformational changes for KaiC monomers in order to obtain oscillations \cite{van2007allosteric, paijmans2017thermodynamically, chang2012rhythmic, zhang2020energy}. However, new structural studies suggest that the positive cooperativity between monomer conformational states is fairly weak \cite{han2023determining}. Hence, it is timely to examine alternative models that can generate emergent oscillations for macromolecules in the presence of strong internal fluctuations or weak positive cooperativity. Overall, our work proposes a new pathway for the emergence of high coherence despite stochasticity and strong fluctuations from the typical paradigm of strong cooperativity, that could be relevant for biological oscillations more generally. 

\section{Topological model for emergent oscillations} \label{sec_model}

We consider discrete stochastic processes that operate in a two-dimensional configuration space. The state of the system is completely specified by three variables $(x,y)_s$. The ``external'' variables $x$ and $y$ are independent dynamical variables. Based on the widespread presence of non-equilibrium cycles in biological systems \cite{hopfield1974kinetic, samoilov2005stochastic,brettschneider2010sequestration}, we propose that the ``external'' transitions modifying these variables form reaction cycles. The external processes have transition rates $\gamma_{\text{ex}}$ and slower reverse rates $\gamma_{\text{ex}}'\ll \gamma_{\text{ex}}$ (black solid arrows), as shown in Fig. \ref{plt_model}a. In the KaiABC system, $x$ and $y$ represent the number of phosphorylated T- and S-sites respectively. T and S are two residues on each monomer of the hexameric KaiC molecule \cite{nishiwaki2004role}, where T phosphorylation is denoted with orange spheres and S phosphorylation with pink spheres in Fig. \ref{plt_model}a. Hence, phosphorylation occurs in the left and top arrows of the cycle in Fig. \ref{plt_model}a, and dephosphorylation in the remaining right and bottom arrows -- this four-state motif contains two futile cycles \cite{hopfield1974kinetic, samoilov2005stochastic}.

The ``internal'' state variable $s$, given by compass directions N-E-S-W, labels which of the four external transitions the system is primed for. For example, in the W state, the system is most likely to go through the westward external transition that decrements $x$. Within each phosphorylation level $(x,y)$, we model transitions between internal states in a cyclic manner, with transition rates $\gamma_{\text{in}}$ and slower reverse rates $\gamma_{\text{in}}'\ll \gamma_{\text{in}}$ (gray dashed arrows in Fig. \ref{plt_model}b). In the KaiABC system, the vertical internal transitions stem from conformational changes of the A-loop \cite{kim2008day,tseng2014cooperative}, where the A-loop is denoted as a blue square when exposed and as a purple circle when buried (Fig. \ref{plt_model}b). The top ($\text{N}\rightarrow \text{W}$) and left ($\text{W}\rightarrow \text{S}$) transitions are facilitated by KaiB binding and unbinding respectively \cite{kitayama2003kaib,snijder2017structures}. The bottom transition ($\text{S}\rightarrow \text{E}$) is catalyzed by KaiA \cite{xu2003cyanobacterial,rust2007ordered}, represented by a curved arrow. As a single KaiA dimer binds and unbinds several times during each phosphorylation \cite{kageyama2006cyanobacterial,mori2018revealing}, it acts as a catalyst, consistent with the treatment of KaiA in previous models \cite{van2007allosteric,lin2014mixtures}. The intermediate KaiA-bound states are coarse-grained out of the model and not shown in Fig. \ref{plt_model}b.

These reactions are repeated for each monomer and hence can be laid out as a lattice. As shown in Fig. \ref{plt_model}c, the phosphorylation/dephosphorylation cycles in Fig. \ref{plt_model}a (green box) and the internal cycles in Fig. \ref{plt_model}b (blue box) both repeat along the $x$ and $y$ axes of T and S phosphorylation. Such a lattice will have edges representing the physical constraints of the system, i.e., $0\leq x \leq N_x$ and $0\leq y \leq N_y$. In our case, $N_x=N_y=6$ since there are 6 sites available on a KaiC hexamer for either T or S phosphorylation. Note that our model only keeps track of the number of T and S phosphorylated monomers, while the specific location of each monomer does not matter.

Our model in Fig. \ref{plt_model}c is mathematically equivalent to the model introduced in \cite{tang2021topology} but given a novel interpretation based on a more realistic description of KaiABC. Specifically, in \cite{tang2021topology} the external variables $x$ and $y$ represent the number of phosphorylated KaiC monomers and the number of monomer conformational changes. In our paper, these variables are interpreted as the number of T and S phosphorylated monomers. Rather than occurring independently, the conformational change is instead hypothesized to change the system internal state. In addition, we include interactions with KaiA and KaiB molecules as well as ADP/ATP turnover as reactions that change the internal state of the model.

Before analyzing the model, we simplify our parametrization for the four transition rates $\gamma_{\text{ex}}$, $\gamma_{\text{ex}}'$, $\gamma_{\text{in}}$, $\gamma_{\text{in}}'$ down to three parameters. First, $\mu$ is the thermodynamic force defined by $e^{\mu/k_B T}\equiv\gamma_{\text{ex}}/\gamma_{\text{ex}}'=\gamma_{\text{in}}/\gamma_{\text{in}}'$. We analyze $\mu$ in units of $k_B T$. For an arbitrary cycle in the network, the sum of $\mu$ along each transition in the cycle is the energy input into that cycle from external driving such as ATP hydrolysis \cite{hill1989}. The system obeys detailed balance when the total $\mu$ along every cycle is 0, and is out of equilibrium otherwise \cite{qian2006open}. Here we assume the same $\mu$ for every transition in the model, which removes one free parameter. Detailed balance then simply corresponds to $\mu=0$. Second, $\rho$ varies the ratio of external to internal transitions, defined by $e^\rho\equiv\gamma_{\text{ex}}/\gamma_{\text{in}}=\gamma_{\text{ex}}'/\gamma_{\text{in}}'$. It quantifies the separation of timescales between the external and internal transitions. Third, $\gamma_{\text{tot}}$ controls the overall timescale of all transitions, i.e., $\gamma_{\text{tot}}\equiv\gamma_{\text{ex}}+\gamma_{\text{ex}}'+\gamma_{\text{in}}+\gamma_{\text{in}}'$.

\begin{figure*}[t]
	\centering
	\includegraphics[width=17cm, height=10.2cm]{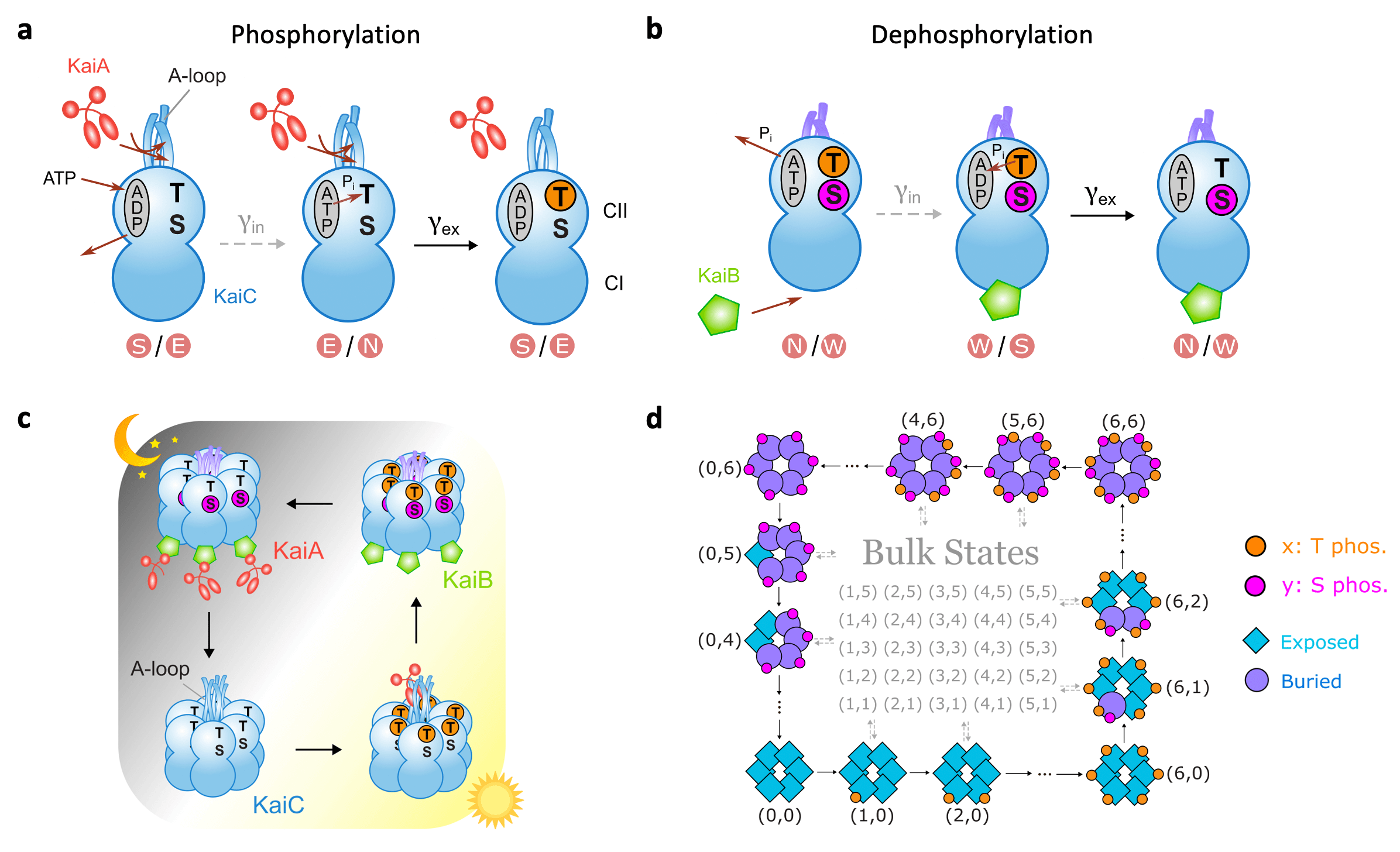}
	\caption{\textbf{Biophysical mechanisms for the topological model, which reproduces the KaiABC circadian rhythm.} \textbf{a}, Phosphorylation of a monomer relies on two main steps. \textit{Left}: Slow reaction $\gamma_{\text{in}}$ where KaiA promotes ADP release and ATP binding in the CII nucleotide binding pocket (gray oval) \cite{nishiwaki2012circadian,egli2012dephosphorylation,nishiwaki2014exchange}, priming the monomer for phosphorylation. \textit{Right}: Fast reaction $\gamma_{\text{ex}}$ where a phosphate group P$_\text{i}$ is transferred from ATP to the phosphorylation site \cite{nishiwaki2012circadian,egli2012dephosphorylation}. \textbf{b,} Dephosphorylation similarly has two main steps. \textit{Left}: Slow reaction $\gamma_{\text{in}}$ where KaiB binds to KaiC \cite{rust2007ordered,abe2015atomic} and ATP hydrolyzes to ADP at CII \cite{nishiwaki2012circadian,egli2012dephosphorylation}, priming the monomer for dephosphorylation. \textit{Right}: Fast reaction $\gamma_{\text{ex}}$ where P$_\text{i}$ is transferred back to ADP \cite{nishiwaki2012circadian,egli2012dephosphorylation}. In both \textbf{a} and \textbf{b}, we illustrate what happens for T-sites; additional distinctions between T and S phosphorylation/dephosphorylation are discussed in the main text. \textbf{c}, KaiABC exhibits oscillations via a concerted global cycle of phosphorylation and dephosphorylation. During the day, all six KaiC monomers get phosphorylated at the T-sites, and then at the S-sites. Phosphorylation is promoted by interaction with KaiA molecules \cite{xu2003cyanobacterial}. By night, phosphorylated KaiC binds to KaiB, which sequesters KaiA from the solution. In the absence of KaiA, all the T-sites get dephosphorylated, followed by the S-sites \cite{rust2007ordered}. Since individual monomers can independently phosphorylate  \cite{brettschneider2010sequestration}, it is unclear why they would perform a concerted phosphorylation cycle that is robust. \textbf{d}, A possible solution lies in the topological phase of the model, in which a global cycle emerges that recapitulates the experimentally observed phosphorylation sequence.}
	\label{plt_KaiC}
\end{figure*}

There are two dynamical regimes for the system, determined by tuning the parameter $\rho$. In the trivial regime when $\rho< 0$, the system tends to go through local counterclockwise cycles via internal transitions $\gamma_{\text{in}}$, interspersed with slower external transitions $\gamma_{\text{ex}}$ that break out of these cycles. Such behavior is observed both in the bulk and on the edge of the state space. Over long times, the system displays diffusive dynamics and will ergodically explore the whole state space. On the other hand, in the topological regime when $\rho>0$, the system supports an edge state \cite{tang2021topology}. In the bulk of the state space, the system would similarly go through local clockwise cycles via external transitions $\gamma_{\text{ex}}$, interspersed with internal transitions $\gamma_{\text{in}}$ which are now slower. However, once the system reaches the edge, it will continue around the edge. This can be verified by inspection in Fig. \ref{plt_model}c (also see Supplementary Movie). Over long times, the steady-state distribution will hence lie on the system edge, forming a global current that flows counterclockwise along the boundary of the lattice (see Fig. \ref{plt_model}d). The edge state and the associated dynamical regime have a topological origin, as their emergence is governed by a topological invariant, the 2D Zak phase \cite{liu2017novel,tang2021topology}. For our topological model, the 2D Zak phase takes the trivial value $(0,0)$ when $\rho<0$ but becomes ${(\pi,\pi)}$ in the topological regime $\rho>0$ (see Supplementary Information for more details). As $\rho$ becomes larger in the topological regime, the Zak phase $(\pi,\pi)$ is preserved, while the edge state becomes more localized on the system boundary \cite{nelson2023non}. The edge state further inherits the useful property of topological protection from inaccessible states \cite{tang2021topology} or perturbations in transition rates (see Supplementary Fig. \ref{plt_SI_robustness} and \ref{plt_SI_disorder}). Because of these unusual properties, we focus on the topological regime and investigate the properties of our system assuming $\rho>0$ from now on.

\section{Biophysical mechanisms for the topological model} \label{sec_experiment}
In the topological regime, our model exhibits a separation of timescales. This finds experimental support in the KaiABC system, where the faster phosphorylation/dephosphorylation reactions are primed by other slower processes. As shown in Fig. \ref{plt_KaiC}a, phosphorylation involves two main steps. The slow transition priming phosphorylation is the KaiA-induced ADP release and ATP binding \cite{nishiwaki2014exchange}. This process occurs in the nucleotide binding pocket on the CII domain \cite{pattanayek2004visualizing} (gray oval on the upper lobe of KaiC in Fig. \ref{plt_KaiC}a). The fast phosphorylation reaction occurs only after the ADP/ATP exchange, as it transfers the phosphate group P$_\text{i}$ from ATP to the phosphorylation site \cite{nishiwaki2012circadian,egli2012dephosphorylation}. Similarly, dephosphorylation also involves two main steps. As shown in Fig. \ref{plt_KaiC}b, the slow transition priming T dephopshorylation is KaiB binding \cite{kageyama2006cyanobacterial,rust2007ordered,abe2015atomic} and ATP hydrolysis at CII \cite{nishiwaki2012circadian,egli2012dephosphorylation}. With ADP at the CII domain, fast dephosphorylation can proceed by transferring P$_\text{i}$ back to ADP \cite{nishiwaki2012circadian,egli2012dephosphorylation}. Such separation of timescales is essential for the emergence of global oscillations in phosphorylation level due to nontrivial topology.

Moreover, non-equilibrium driving powers key reactions in the KaiABC phosphorylation cycle. In particular, phosphorylation is powered by conversion of ATP to ADP \cite{nishiwaki2012circadian,egli2012dephosphorylation}, a process that dissipates free energy in relevant experimental conditions. Indeed, experiments show that the ATP consumption rate increases by 75\% during the phosphorylation phase compared to the average basal rate \cite{terauchi2007atpase}. In addition, KaiB binding is powered by ATP hydrolysis in the CI domain (lower lobe of KaiC in Fig. \ref{plt_KaiC}), which leads to subsequent dephosphorylation \cite{phong2013robust}. CI ATP hydrolysis proceeds slowly but continuously throughout the phosphorylation cycle, at a rate of only $\sim 10$ ATP molecules per day \cite{terauchi2007atpase,phong2013robust}. Such consistent external driving is required for a global reaction cycle where each transition is biased toward the same direction and detailed balance is broken \cite{hill1989}.

In Fig. \ref{plt_KaiC}a and \ref{plt_KaiC}b, we illustrate the case of T phosphorylation/dephosphorylation. The same ADP/ATP exchange and phosphotransfer mechanisms are used for S phosphorylation/dephosphorylation. Additionally, S phosphorylation is accompanied by A-loop burial while S dephosphorylation is accompanied by A-loop exposure in our model, consistent with experimental results \cite{tseng2014cooperative}. In particular, we assume that monomer conformational changes occur after S phosphorylation of the same monomer. It follows that the states $(x,y)_\text{E}$ and $(x,y)_\text{S}$ (for $y>0$) have $y-1$ circles and $7-y$ squares, while the states $(x,y)_\text{W}$ and $(x,y)_\text{N}$ have $y$ circles and $6-y$ squares. Moreover, KaiB unbinding rather than KaiB binding tends to occur during S dephosphorylation \cite{snijder2017structures}. We note this trend by ``-KaiB'' in Fig. \ref{plt_model}b. Note that the internal processes illustrated in Fig. \ref{plt_KaiC}a and \ref{plt_KaiC}b form reaction cycles as shown in Fig. \ref{plt_model}b. This is consistent with experimental evidence that A-loop exposure facilitates KaiA catalysis \cite{kim2008day} and inhibits KaiB binding \cite{chang2011flexibility}, while A-loop burial facilitates KaiB binding and inhibits KaiA catalysis \cite{chang2011flexibility}. Such interactions between A-loop conformations and KaiA/KaiB therefore restrict the allowed internal transitions to a single cyclic pathway.

In experimental conditions that support oscillations, KaiC forms a stable hexameric structure \cite{mori2002circadian}. Each monomer in the hexamer can independently go through the biochemical reactions illustrated in Figure \ref{plt_KaiC}a and \ref{plt_KaiC}b. Experimentally, it is well established that KaiC molecules exhibits 24-hour cycles via a concerted phosphorylation sequence in the presence of KaiA and KaiB \cite{nakajima2005reconstitution}. As illustrated in Fig. \ref{plt_KaiC}c, during the day, all six KaiC monomers get phosphorylated at the T-sites, and then at the S-sites. By night, all the T-sites get dephosphorylated, again followed by the S-sites \cite{cohen2015circadian}. The phosphorylation phase is facilitated by KaiA catalysis \cite{xu2003cyanobacterial}, while the dephosphorylation phase is promoted by KaiB binding \cite{kitayama2003kaib}. The robustness of this concerted phosphorylation sequence is surprising, since KaiC monomers within the hexamer can independently phosphorylate and change conformations \cite{han2023determining}. Given this very large phase space of possible reactions available to individual monomers, why would all the monomers phosphorylate together in a concerted global cycle? Further, since phosphorylation in both T and S are promoted during the day, why does phosphorylation proceed in the specific order where all six T-sites get phosphorylated before S phosphorylation begins (Fig. \ref{plt_KaiC}c)? 

To answer this question, models \cite{van2007allosteric, paijmans2017thermodynamically, chang2012rhythmic, zhang2020energy} have typically relied on the concerted or Monod-Wyman-Changeux paradigm \cite{monod1965nature} of allosteric regulation, which restricts the configuration space such that either all or none of the monomers in a complex undergo conformational change. However, recent cryo-EM data shows that monomers can demonstrate independent conformational changes in the same hexamer \cite{han2023determining}, challenging this strong restriction. Further, these models often put in by hand the specific ordering of T phosphorylation occurring before S phosphorylation.

Our topological model presents an alternative way to account for the experimental facts and explain the emergence of the KaiC phosphorylation cycle. As discussed in Sec. \ref{sec_model}, in the topological regime where $\rho\gg0$, the system supports a propagating edge current. Fig. \ref{plt_KaiC}d shows a coarse-grained picture of the edge current. One out of four internal states is shown for each phosphorylation level $(x,y)$ along the edge, specifically the last state along the edge (e.g. E states for the bottom edge or S states for the left edge). As we can see, the edge current is equivalent to a global cycle of concerted phosphorylation of the T-sites, followed by the S-sites, then dephosphorylation of the T-sites and, lastly, of the S-sites (also see Supplementary Movie). This provides a mechanism that allows for individual monomers to undergo conformational and other changes, while still producing a global cycle and the experimentally observed phosphorylation sequence that emerges with less fine-tuning.

\vspace{10mm}

\section{Model thermodynamics: Precision vs cost} \label{sec_properties}

While this topological model provides an unique alternative mechanism to experimentally observed oscillatory dynamics, how precise or efficient are the oscillations produced? In this section, we quantify the thermodynamics and entropy production of this model, and compare its performance to that of other KaiC models \cite{barato2017coherence, marsland2019thermodynamic, li2009circadian}. Further, we identify a new predictor for oscillator coherence and analyze the saturation of thermodynamic bounds for different models. We begin by analyzing the master equation that describes stochastic systems, 
\begin{equation} \label{eqn_master_eqn}
    \frac{d\bp}{dt}=\cW\bp,
\end{equation}
where $\bp(t)$ is a vector that describes the probability distribution of system states. $\cW$ is the transition matrix, whose elements $\cW_{ij}$ specify the transition rates from state $j$ to $i$. The dynamics of oscillations is typically dominated by the first non-zero eigenvalue of $\cW$, which is the eigenvalue with the smallest modulus in the real part \cite{barato2017coherence}. This eigenvalue is denoted as $-\lambda_\text{R} \pm i\lambda_\text{I}$ for the real and imaginary parts $\lambda_\text{R}$ and $\lambda_\text{I}$, respectively. Here $\lambda_\text{R}, \lambda_\text{I}\geq 0$ and we take $\lambda_1=-\lambda_\text{R} + i\lambda_\text{I}$. In general, $\bp(t)$ relaxes to the steady-state distribution through damped oscillations, with a decay time $\lambda_\text{R}^{-1}$ and oscillation period $\mathcal{T}=2\pi/\lambda_\text{I}$ \cite{barato2017coherence} (also see Supplementary Fig. \ref{plt_SI_corrfunc}). Following \cite{barato2017coherence}, we define coherence as the ratio
\begin{equation} \label{eqn_cohr}
    \cR \equiv \frac{\lambda_\text{I}}{\lambda_\text{R}},
\end{equation}
which quantifies the robustness of sustained oscillations before stochastic noise destroys the coherence (more details in Supplementary Information). We would like to see how our model performs using this metric and what factors contribute to increased coherence. 

\begin{figure*}[]
	\centering
	\includegraphics[width=17cm, height=10.4cm]{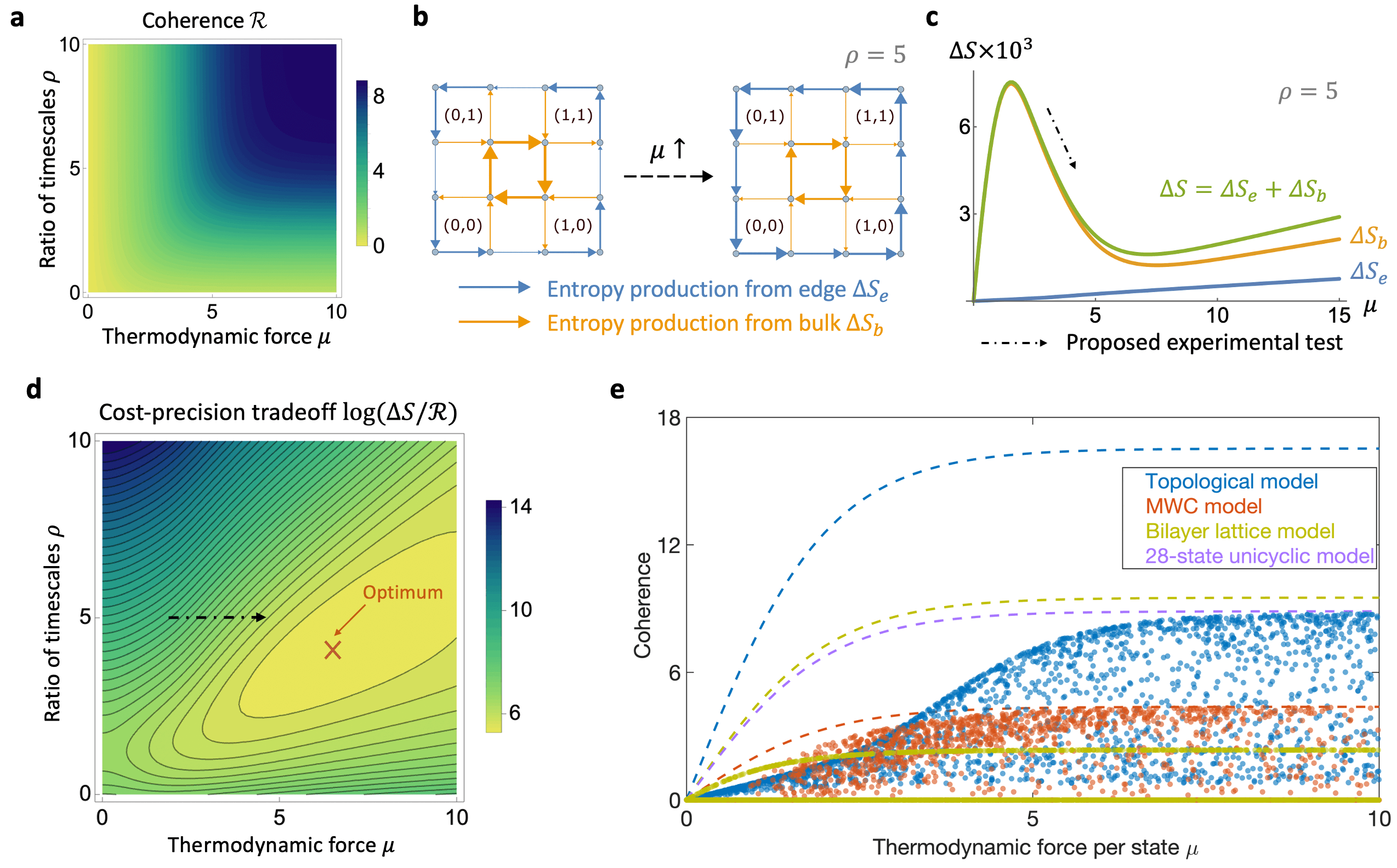}
	\caption{\textbf{Coherence and an efficient regime with simultaneously increased precision and decreased cost.} \textbf{a}, Phase diagram of coherence $\cR$ for the topological model, which increases with respect to thermodynamic force $\mu$ and ratio of timescales $\rho$ as expected. \textbf{b}, Entropy production per period $\Delta S$ moves from the bulk to the boundary of the system with increasing $\mu$, illustrated on a smaller lattice. The system is in the topological regime; $\rho=5$. Blue arrows represent entropy production on the edge $\Delta S_\text{e}$ and orange arrows represent entropy production in the bulk $\Delta S_\text{b}$. Arrow thickness corresponds to the magnitude of the entropy production for the corresponding transitions and arrows point in the direction of the probability flux. \textbf{c}, Entropy production per period $\Delta S$ for the topological model ($N_x=N_y=6)$ is decomposed into a bulk and an edge contribution. Due to the localization effects in Fig. \ref{plt_coherence}b, the entropy production on the edge $\Delta S_\text{e}$ (blue) increases while entropy production in the bulk $\Delta S_\text{b}$ (orange) decreases with $\mu$ around $1.5<\mu<7$. Typically, the bulk contribution dominates the edge contribution, hence their sum $\Delta S$ (green) also decreases with increasing $\mu$. Experimentally, the negative slope in this regime predicts a decreased overall ATP consumption when ATP concentration is increased, as indicated by the black dash-dotted arrow. \textbf{d}, Phase diagram of the cost-precision tradeoff $\Delta S/\cR$ \cite{oberreiter2022universal}. The black dash-dotted arrow, similar to that in \textbf{c}, indicates increasing $\mu$ in the energy-efficient regime. This leads to a lower $\Delta S/\cR$, which implies a more cost-effective oscillator. There is a global optimum for $\Delta S/\cR$ at $\mu=6.5, \rho=4.1$. \textbf{e}, Comparison of coherence for different KaiC models with randomly sampled parameters (individual points): there is a strongly-driven regime where the topological model has the highest coherence. The dashed lines represent the upper bounds on coherence for the corresponding models in terms of the thermodynamic force $\mu$ \cite{barato2017coherence} (see Methods). The purple dashed line is the upper bound for the 28-state unicyclic model, which is approached by the edge state in the strongly topological regime. Values of $\Delta S$ in all panels are given in units of $\gamma_{\text{tot}}k_B $.}
	\label{plt_coherence}
\end{figure*}

In typical oscillator models, coherence can be increased by dissipating more free energy. For a general oscillator model described by a master equation, the free energy cost for maintaining the non-equilibrium steady state under constant temperature can be quantified by the entropy production per period $\Delta S$. Denoting the steady-state probability distribution as $\bp^s$ and the oscillation period as $\mathcal{T}$, $\Delta S$ is given by \cite{schnakenberg1976network,ge2010physical}
\begin{equation}
    \Delta S = \frac{\mathcal{T}}{2}\sum\limits_{i,j}(p_j^s \cW_{ij} - p_i^s \cW_{ji})\ln{\left( \frac{p_j^s \cW_{ij}}{p_i^s \cW_{ji}} \right)}.
\end{equation}
In the MWC-type model of KaiC studied in \cite{barato2017coherence}, for example, increasing $\Delta S$ is necessary to increase coherence. Both quantities increase when the external driving $\mu$ is stronger, although coherence starts to decrease when $\mu$ is increased still further (see Supplementary Fig. \ref{plt_SI_cost_precision}). This leads to even worse performance for the oscillator as it maintains less coherent oscillations with increasing energetic cost.

On the contrary, our model in the topological regime displays an unusual regime where coherence increases while entropy production per period becomes lower. Fig. \ref{plt_coherence}a shows the coherence of our model as a function of the two parameters $\mu$ and $\rho$ (also see Supplementary Fig. \ref{plt_SI_cohr} featuring both positive and negative $\rho$). Increasing the thermodynamic force increases coherence monotonically, as expected for typical oscillator models \cite{barato2017coherence}. Going deeper into the topological regime by increasing $\rho$ also leads to higher coherence, as the global currents become more localized on the system edge \cite{nelson2023non}. However, the entropy production per period $\Delta S$ does not change monotonically with $\mu$. As illustrated in Fig. \ref{plt_coherence}b in a smaller lattice, the system response becomes localized to the edge as $\mu$ increases. This causes the entropy production on the edge (blue arrows) to increase, while the entropy production in the bulk (orange arrows) decreases, in the region $1.5<\mu<7$. Since the bulk contribution typically dominates the edge contribution, i.e., $\mathcal{O}(N^2)\gg \mathcal{O}(N)$ where $N$ is the typical system size, the sum of their contributions also decreases (green curve in Fig. \ref{plt_coherence}c). This negative slope of $\Delta S$ with respect to $\mu$ implies that the system dissipates less free energy overall even when the external driving $\mu$ supplied to each reaction is stronger. In other words, we ``get more from pushing less'' \cite{zia2002getting,conwell1970negative,kostur2006forcing,jack2008negative}. This unusual regime has a topological origin, since the decrease in the bulk entropy production results from the localization of the steady state onto the system edge. This unusual negative slope leads to a unique experimental signature. Where $\Delta S$ has a negative slope, increasing the ATP concentration (increasing $\mu$) leads to a decrease in total ADP production (decreasing $\Delta S$). This is a striking prediction of our topological model, as shown by the dash-dotted arrow in Fig. \ref{plt_coherence}c. In addition, since coherence increases monotonically with $\mu$, this leads to an efficient regime with simultaneously increasing coherence and decreasing cost. The cost-effectiveness in terms of expending free energy for coherent oscillations can be measured by the ratio $\Delta S/\cR$ \cite{oberreiter2022universal}. See Fig. \ref{plt_coherence}d for the region in parameter space with low $\Delta S/\cR$, indicating a highly efficient oscillator.

We can further compare coherence between different families of KaiC models. We include a simple MWC-type model \cite{barato2017coherence} and a bilayer model that has a lattice structure more similar to ours, adapted from Li et al. \cite{li2009circadian} (details of each in Supplementary Information). Each layer in the bilayer lattice represents the T and S phosphorylation levels along its $x$ and $y$ coordinates similar to our topological model, with the possibility to switch between the two layers that denote unbound KaiC and KaiB-bound KaiC, respectively (See Supplementary Fig. \ref{plt_SI_model}b and \ref{plt_SI_model}c). KaiC is more likely to bind to KaiB on the upper right half of each lattice and more likely to unbind on the lower left half. In order to aid comparison, we simplify this lattice model using our $(\mu,\rho)$ parameters (details in Supplementary Information).

By sampling random parameters in the three models, we find a regime of high thermodynamic driving where the topological model has the highest coherence (Fig. \ref{plt_coherence}e).  In the same plot, we also indicate a thermodynamic bound for coherence \cite{barato2017coherence} for each model (dashed lines), which depend on the system size and thermodynamic force $\mu$ (see Methods). We also plot the bound for a 28-state unicyclic model (purple dashed line in Fig. \ref{plt_coherence}e), which our model approaches in the limit of high $\mu$ and high $\rho$. This is because there are 28 slow internal transitions on the edge that form the effective bottleneck and dominate over the other fast external transitions, resulting in 28 high-probability states as noted in Fig. \ref{plt_model}d. This shows that our topological model approaches the bound set by the most coherent cycle, which is the unicycle with uniform rates ($\gamma_{\text{in}}$ and $\gamma_{\text{in}}'$ in our case) \cite{barato2017coherence}, deep in the topological regime with high external driving.

\begin{figure}[]
	\centering
	\includegraphics[width=9cm, height=10.5cm]{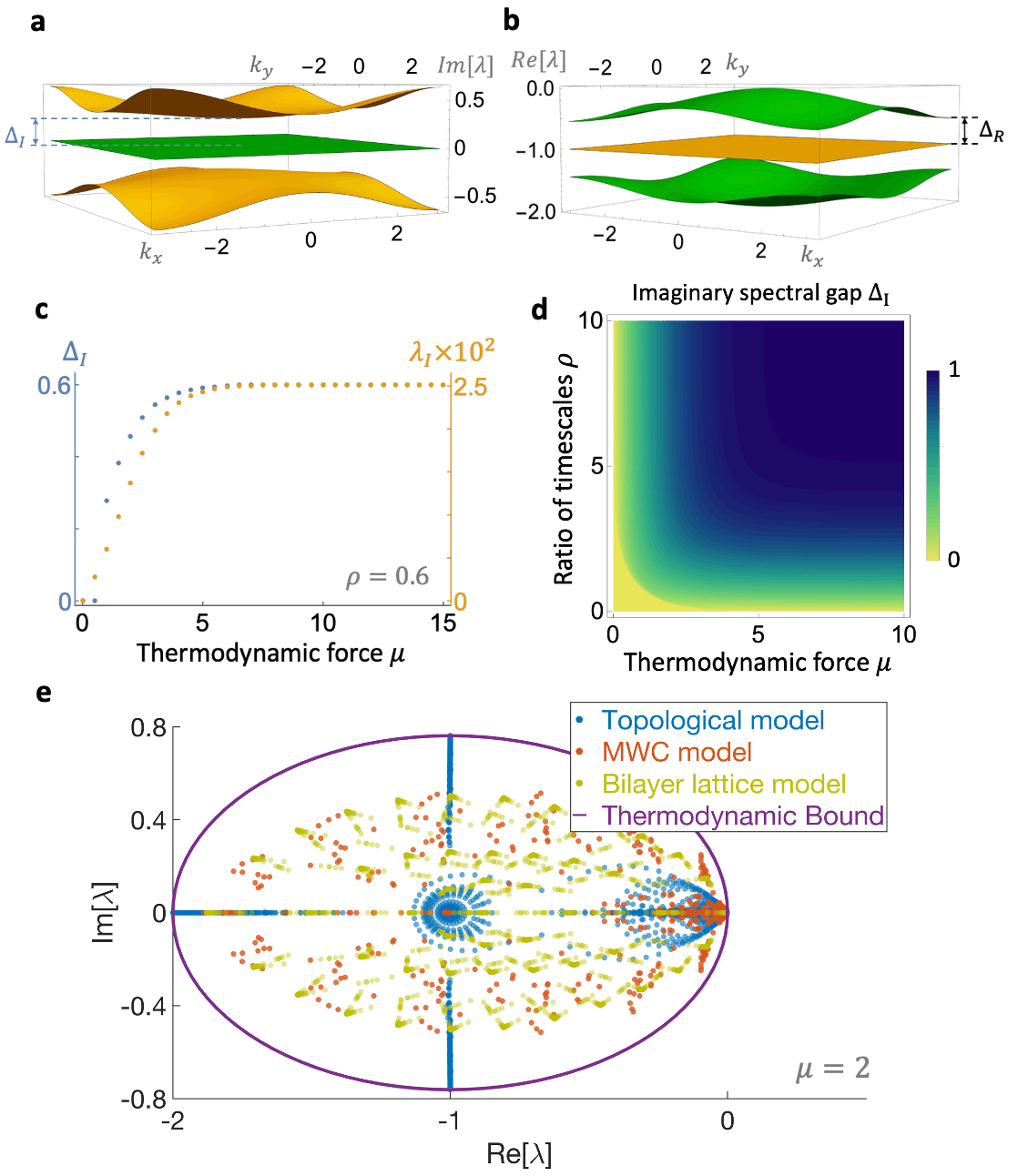}
    	\caption{\textbf{New predictor of coherence and saturation of a global thermodynamic bound.} \textbf{a},\textbf{b}, Spectrum for the topological model in reciprocal space in imaginary space (\textbf{a}) and real space (\textbf{b}) for $\mu=1.5$ and $\rho=1$. The spectral gaps $\Delta_\text{I}$ and $\Delta_\text{R}$ are defined as the difference between the minimum of the topmost band and the maximum of the second band. \textbf{c}, The imaginary spectral gap $\Delta_\text{I}$ closely tracks the imaginary part $\lambda_\text{I}$ of the first non-zero eigenvalue $\lambda_1$, as a function of $\mu$. \textbf{d}, Phase diagram for the imaginary spectral gap $\Delta_\text{I}$ in $\mu$ and $\rho$. $\Delta_\text{I}$ and coherence (Fig. \ref{plt_coherence}a) track each other monotonically, and both saturate to a maximum value after a rapid increase. \textbf{e}, The global spectral bound for a fixed $\mu$ is shown by a purple ellipse in the complex plane \cite{uhl2019affinity}, while points represent the spectra for the transition matrix $\cW$ for different oscillator models. The topological model approaches this bound in the limit $\rho\rightarrow\infty$, which is saturated by the most coherent cycle (the unicycle) \cite{uhl2019affinity}. Here $\mu=2$ and other parameters are randomly sampled. In all panels, eigenvalues are given in units of $\gamma_{\text{tot}}$.}
	\label{plt_cohr_gap}
\end{figure}

Given the high coherence of our model, we would like to identify the factors that determine high coherence. Here, we introduce a new predictor of coherence, which is the spectral gap (or band gap) of the system, inspired from band theory of solids \cite{kittel1955solid}. As the spectral gap measures the separation between modes with different timescales \cite{schnakenberg1976network}, a larger gap predicts greater separation between modes and hence the stability of longer-lived modes, as they mix less with transient ones. See Fig. \ref{plt_cohr_gap}a and \ref{plt_cohr_gap}b for the spectral gap in both imaginary and real space, $\Delta_\text{I}$ and $\Delta_\text{R}$ respectively, in the reciprocal space version of our topological model (see Supplementary Information for more details). The spectral gap is defined as the difference between the minimum of the topmost band and the maximum of the band below it. Comparing $\Delta_\text{I}$ with the imaginary part $\lambda_\text{I}$ of the first non-zero eigenvalue $\lambda_1$ in Fig. \ref{plt_cohr_gap}c, we see that they track each other well as $\mu$ is increased. Since coherence $\cR$ is the ratio of $\lambda_\text{I}$ to its corresponding real part $\lambda_\text{R}$ of the same eigenvalue (Eq. \ref{eqn_cohr}) and $\lambda_\text{R}$ remains roughly constant with increasing $\mu$ (see Supplementary Fig. \ref{plt_SI_lambda_cohr}), $\Delta_\text{I}$ tracks $\cR$ (Fig. \ref{plt_coherence}a) monotonically in $\mu$ as well. Along the other axis, increasing $\rho$ decreases the spectral dispersion, i.e. it compresses the top band in Fig. \ref{plt_cohr_gap}a. The topmost point of the top band also increases with $\rho$ while the middle green bands in Fig. \ref{plt_cohr_gap}a remain flat, leading to a widened spectral gap $\Delta_\text{I}$ for increasing $\rho$. Moreover, $\lambda_\text{R}$ decreases more quickly than $\lambda_\text{I}$ with increasing $\rho$, so that their ratio $\cR$ increases (see Supplementary Fig. \ref{plt_SI_lambda_cohr}). Therefore, $\Delta_\text{I}$ tracks coherence in $\rho$ as well. Hence, both the imaginary spectral gap and coherence track each other monotonically, as can be seen in their phase diagrams Figs. \ref{plt_cohr_gap}d and \ref{plt_coherence}a respectively, where both saturate to a maximum following a rapid increase.

Lastly, we examine how the different models perform compared to a conjectured global thermodynamic bound on the full spectrum of $\cW$ \cite{uhl2019affinity}. For a given thermodynamic force, the spectrum is found to lie within an ellipse in the complex plane (purple line in Fig. \ref{plt_cohr_gap}e). The most coherent cycle, a unicyclic network with uniform rates \cite{barato2017coherence}, saturates this bound \cite{uhl2019affinity}. While the other two families of KaiC models do not saturate the bound for any sampled parameters, our model approaches this bound as $\rho\rightarrow\infty$. This is consistent with our model approaching a uniform unicyclic network when $\rho$ is large, and also with our previous analysis of the spectral gap contributing to high coherence. Similar to the results in reciprocal space model discussed above, the transition matrix $\cW$ also shows a larger imaginary spectral gap as $\rho$ increases. The larger gap moves the topmost eigenvalue upward in the complex plane to saturate the global spectral bound while increasing coherence (see Supplementary Information and Fig. \ref{plt_SI_OBC} for details). 

\section{Discussion} \label{sec_discussion}

We have proposed a topological model that generates coherent oscillations, which supports an unusual regime with increased coherence and simultaneously decreased energetic cost. The mathematical model, which first appeared in \cite{tang2021topology}, is interpreted in a biological context and applied to the KaiABC system. Compared to other KaiABC models, our model has high coherence and more closely saturates a global spectral bound, similar to the most coherent unicyclic models. We also find that the imaginary spectral gap can be used to predict oscillation coherence. Further, the kinetic ordering of the KaiABC phosphorylation cycle arises naturally as an edge current in our model. 

In contrast to typical MWC-type models which usually involve a large number of system-dependent parameters \cite{van2007allosteric,paijmans2017thermodynamically}, our model is parsimonious in that it captures the same phosphorylation dynamics with only a few parameters. Moreover, it also does not require fine-tuning of the reaction rates or restriction of the configuration space to all-or-none conformational change, an assumption that is not supported by emerging experimental evidence \cite{han2023determining}. Indeed, recent cryo-EM images of KaiC hexamers suggest that KaiC conformational changes are in a weak coupling regime \cite{han2023determining}. From experimental data, the two states consistent with the MWC picture (all buried or all exposed A-loops) are not the most abundant species observed. Rather, hexamers with mixed A-loop conformations make up more than 80\% of the entire population. Notably, the coupling constant $J$ from the Ising model fit is very small ($J=0.086$) and the correlation length $\xi$ is short ($\xi_{AA}\approx \xi_{EE}\approx0.5$ monomer length) -- demonstrating weak cooperativity between monomers.

We further propose three testable experimental signatures for our topological model in the KaiABC system. First, we predict a response unique to our topological model where decreasing ATP concentration leads to an increase in ADP consumption. This is the opposite of what would typically be expected and is due to the system being driven from the edge localized state into bulk dynamics, that simultaneously increases energy consumption despite a decrease in coherence. This is unique to our topological model and discussed in Fig. 3c. We expect a 5-fold increase in ADP consumption for a 10-fold decrease in ATP concentration, which can be measured via tracking ADP levels \cite{terauchi2007atpase} or alternatively with heat dissipation \cite{bae2021micromachined}. Second, we predict that KaiC hexamers should have a wide distribution of different A-loop conformational patterns when imaged over the period of an oscillation. This would be a similar pattern to what has been observed for non-oscillating KaiC mutants \cite{han2023determining}, which would distinguish our model from MWC-type models \cite{van2007allosteric,paijmans2017thermodynamically,lin2014mixtures}, where all six A-loops are expected to be mostly buried or mostly exposed. Lastly, mutants that do not oscillate \cite{nishiwaki2007sequential} can be used to avoid potential confounds due to synchronization effects. As our model couples S phosphorylation with A-loop burial (while still allowing independent conformational change), this predicts that mutants mimicking phosphorylated S-sites would correlate with hexamers having mostly buried A-loops with the highest frequency. Meanwhile, mutants mimicking unphosphorylated S-sites would correlate with hexamers having mostly exposed A-loops.

Our model gives rise to the observed sequence of phosphorylation reactions in KaiABC without having to tune as many free parameters as in typical models \cite{van2007allosteric, paijmans2017thermodynamically}. This is due to the repetition of simple motifs in our network. Further, we note that population effects in KaiABC are known to be important, e.g., by promoting the synchronization of many molecules in the KaiABC system \cite{nakajima2005reconstitution, rust2007ordered, van2007allosteric}. Still, the oscillation and coherence of individual molecules are generally presumed to be building blocks for sustained oscillations at the population level, even while the extent of the single-molecule contribution remains unclear. Hence, our work focuses on the oscillations and coherence of single molecules as a first step towards the understanding of population-level coherence more generally, with the exploration of multiple molecules and their synchronization to be left for future work. See Section \ref{sec_many_molecules} in Supplementary Information for preliminary work on generalizing our single-molecule model to many molecules and incorporating population-level effects such as competition for KaiA molecules.

By rigorously embedding topological methods within non-equilibrium statistical physics, our work generalizes their usage for various biological and chemical systems. Our results suggest a new mechanism that utilizes dissipative cycles to produce emergent oscillations or attractor states in biological systems. This mechanism can be tested by introducing perturbation or mutant proteins \cite{chavan2021reconstitution}, in order to jointly analyze how these changes modify the robustness and coherence of the global cycle. Even though we have mostly discussed our model in the context of the KaiABC system, the model is more general. It can be mapped to various other biological systems where $x$ and $y$ represent other possible types of molecular modifications such as polymerization \cite{tang2021topology}. More broadly, our model provides a blueprint for the design for synthetic oscillators, which is becoming increasingly feasible due to new experimental developments \cite{klingel2022model, chen2015transplantability}. While designing synthetic oscillators that are robust across different parameters or changes in the environment remains a challenge \cite{woods2016statistical, puvsnik2019computational}, this project provides new models for robust oscillators and continuous attractor dynamics in various biochemical scenarios and changing conditions.

\section*{Data Availability}

Data will be openly accessible once the paper is published.

\section*{Code Availability}

Code will be openly accessible once the paper is published.

\bibliography{references}

\balancecolsandclearpage

\section*{Methods}

\textbf{Simulation of the system steady-state.} The steady-state probability distribution in Fig. \ref{plt_model}d is obtained by simulating the system dynamics with the Gillespie algorithm \cite{gillespie1977exact}. The simulation is run for $10^8$ steps with a random initial condition, and the probability for each state is given by the fraction of time the system spends in that state. By calculating the steady-state probability flux $J_{ij}=p^s_j \cW_{ij} - p^s_i \cW_{ji}$ from state $j$ to $i$, we obtain the global counterclockwise current along the edge of the lattice, illustrated with black arrows in Fig. \ref{plt_model}d.

\textbf{Thermodynamic bound for coherence.} Ref. \cite{barato2017coherence} conjectures an upper bound for coherence for any stochastic system. Suppose that a cycle $\kappa$ has $N_\kappa$ states labeled by $\kappa_1,\kappa_2,...,\kappa_{N_\kappa}$ such that $\kappa_1$ is connected to $\kappa_{N_\kappa}$ and $\kappa_2$, $\kappa_2$ is connected to $\kappa_1$ and $\kappa_3$, etc. Define the affinity of $\kappa$ as $\mathcal{A_\kappa} \equiv \ln{\prod\limits_{i=1}^{N_\kappa} \frac{\cW_{\kappa_{i+1}\kappa_{i}}}{\cW_{\kappa_{i}\kappa_{i+1}}}}$, where $\kappa_{N_\kappa+1}$ is the same as $\kappa_1$. For an arbitrary stochastic model, we look at all possible cycles in the underlying network. The upper bound for coherence in Fig. \ref{plt_coherence}e is given by 
\begin{equation*}
    \cR\leq\max\limits_\kappa\left\{\cot{(\pi/N_\kappa)} \tanh{[\mathcal{A_\kappa}/(2N_\kappa)]}\right\}.
\end{equation*}
For the topological model, the thermodynamic force per state $\mathcal{A_\kappa}/N_\kappa$ is always $\mu$, and the cycle that maximizes the right hand side of the bound is the global cycle going around the boundary with $N_{\kappa}=52$.

\textbf{Global spectral bound for a driven stochastic system.} The global bound in Fig. \ref{plt_cohr_gap}e is conjectured by \cite{uhl2019affinity} for the spectrum of any transition matrix for a master equation. To obtain the bound, we look for a cycle that maximizes the affinity per state $\mathcal{A_\kappa}/N_\kappa$. Denote this maximum by $\mathcal{A}_\mathcal{C}/N_\mathcal{C}$. We also define $w_0=\max\limits_i[\bigl|\cW_{ii}\bigr|]$. The spectrum is hypothesized to lie entirely in the ellipse given by
\begin{align*} \label{eqn_spectrum_bound}
\begin{split}
    g(x) = w_0 \bigl\{-&1+\cos{(2\pi x)} \\ 
    + &i\tanh{[\mathcal{A}_\mathcal{C}/(2N_\mathcal{C})]}\sin{(2\pi x)} \bigr\}.
\end{split}
\end{align*}
For the topological model, we have $\mathcal{A}_\mathcal{C}/N_\mathcal{C}=\mu$. The bound for $\mu=2$ is plotted in purple in Fig. \ref{plt_cohr_gap}e. Each model has a different $w_0$ for their corresponding transition matrix. To plot all spectra under a common bound, we rescale each transition matrix by a constant factor such that $w_0=1$ for all transition matrices considered.

\textbf{Sampling random parameters} In Fig. \ref{plt_coherence}e, for the topological model and the bilayer lattice model, we randomly select the parameters $\mu\in[0,10]$ and $\rho\in[0,7]$ from uniform distributions on each interval. For the MWC model in \cite{barato2017coherence} (see Supplementary Information for parameter definitions), we select from the uniform distributions $\gamma\in[3,7]$, $E\in[5,15]$ and $\eta\in[0,\frac{70}{3}]$. The parametrization of the MWC model is such that the maximum affinity per state is $\mathcal{A}/N=\frac{3}{7}\eta$. Because we keep $\mathcal{A}/N$ (which we also call thermodynamic force per state and simply denote as $\mu$) the same in Fig. \ref{plt_coherence}e, $\eta\in[0,\frac{70}{3}]$ exactly corresponds to $\mu\in[0,10]$. For Fig. \ref{plt_cohr_gap}e, we fix the thermodynamic force per state $\mu=2$ (which is $\eta=\frac{14}{3}$ for the MWC model) and sample the remaining parameters $\rho$, $\gamma$, and $E$ in the same way as above. 

\section*{Acknowledgement}
We are grateful to Jordan Horowitz, David Lubensky, Jaime Agudo-Canalejo, Yuhai Tu and Michael Rust for helpful discussions. In addition, we thank Pankaj Mehta, Peter Wolynes, Ulrich Schwarz and Oleg Igoshin for their thoughtful comments. C.Z. and E.T. acknowledge support from NSF CAREER Award (DMR-2238667) and E.T. acknowledges support from the NSF Center for Theoretical Biological Physics (PHY-2019745).

\end{document}


\preprint{APS/123-QED}

\title{Supplementary Information for ``A topological mechanism for robust and efficient global oscillations in biological networks"}

\author{Chongbin Zheng}
\affiliation{
 Department of Physics and Astronomy, Rice University, Houston, Texas 77005, USA
}
\affiliation{
 Center for Theoretical Biological Physics, Rice University, Houston, Texas 77005, USA
}
\author{Evelyn Tang}
\affiliation{
 Department of Physics and Astronomy, Rice University, Houston, Texas 77005, USA
}
\affiliation{
 Center for Theoretical Biological Physics, Rice University, Houston, Texas 77005, USA
}

\maketitle

\onecolumngrid

\section{Robustness of oscillations in our model \label{sec_topology}}

\begin{figure}[H]
	\centering
	\includegraphics[width=15cm, height=8.6cm]{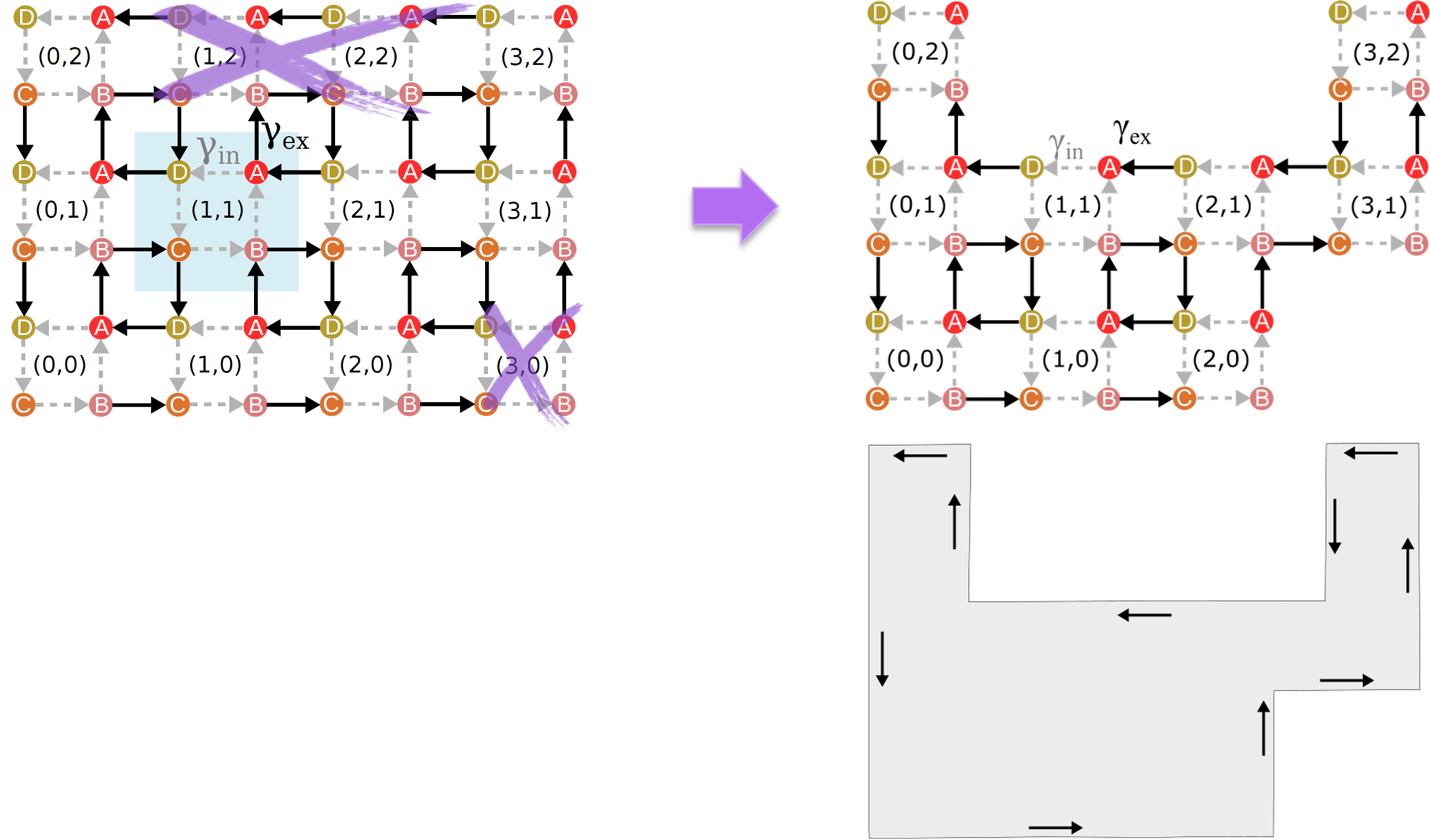}
	\caption{\textbf{Topological protection ensures robustness of the edge state to obstacles or missing components.} In the presence of missing components or obstacles (purple crosses), the edge state (when $\gamma_{\text{ex}}\gg\gamma_{\text{in}}$) will simply go around them to maintain the largest available phase space. This robustness of the edge state can shed light on how biological systems can flexibly pivot in the presence of changing conditions or external stimuli. For illustration purposes, we use a smaller system with $N_x=4,N_y=3$ and only show the forward transitions $\gamma_{\text{ex}}$ and $\gamma_{\text{in}}$.}
	\label{plt_SI_robustness}
\end{figure}

In the topological regime ($\rho\gg 0$), our model supports global currents that propagate along the edge of the state space. The edge currents turn out to be robust against changes in the environment that render certain states inaccessible. For example, a limited number of KaiA molecules in solution could prevent the system from accessing the highly phosphorylated states, while a limited number of KaiB could block off hypophosphoryalated states where $x$ and $y$ are small. Despite missing certain states in the lattice, the probability currents in our model can continue to propagate along the new edge of the state space, as illustrated by Fig. \ref{plt_SI_robustness}. The robustness in edge currents could explain how biological systems maintain stable dynamics in the face of changing external conditions.

\begin{figure*}[]
	\centering
	\includegraphics[width=18cm, height=4.5cm]{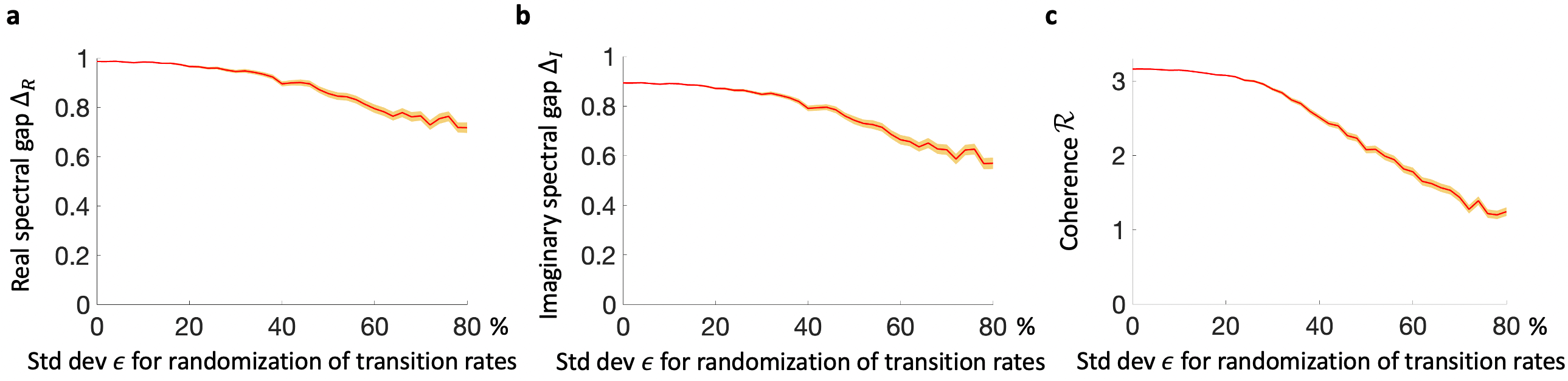}
	\caption{\textbf{Spectral gaps and coherence for randomized transition rates.} \textbf{a}, Each transition rate ($\gamma_{\text{ex}},\gamma_{\text{ex}}',\gamma_{\text{in}},\gamma_{\text{in}}'$) in the model is multiplied by a different random scaling factor $f$ in each direction (N,S,E,W). $f$ is taken from a normal distribution with mean $1$ and standard deviation $\epsilon$. In the case when a negative value is sampled, we take $f=0$. Here we plot the real spectral gap $\Delta_\text{R}$, averaged over different random realizations of transition rates, as a function of the standard deviation $\epsilon$ for randomization of the transition rates. The randomized rates repeat along the $x$ and $y$ axis. For example, any eastward $\gamma_{\text{ex}}$ transition has the same rate after randomization regardless of the $(x,y)$ coordinates. \textbf{b}, Average imaginary spectral gap $\Delta_\text{I}$ as a function of $\epsilon$. \textbf{c}, Average coherence $\cR$ as a function of $\epsilon$. For all panels, we average over 500 random configurations for each $\epsilon$. The same configurations are used to calculate all three quantities. The yellow shaded area represent one standard error. The initial parameters for the model before multiplying by $f$ are $\mu=3,\rho=5$.}
	\label{plt_SI_disorder}
\end{figure*}

The global oscillations in our model are also robust to perturbations in transition rates. The changes in rates can come from changing external conditions like concentrations of KaiA and KaiB or setting more biologically realistic transition rates for each reaction. They can also come from incorporation of additional KaiC reactions hitherto unconsidered. For instance, ATP association with the nucleotide binding site plays the same role as KaiA interaction in promoting T phosphorylation \cite{nishiwaki2012circadian}. This reaction can be combined into the $S\rightarrow E$ internal transition, which results in an effective rate for the combined reaction that is larger than the rates for either reactions. To study the effects of such modifications, we make the rates non-uniform in different directions by multiplying each of $\gamma_{\text{ex}},\gamma_{\text{ex}}',\gamma_{\text{in}},\gamma_{\text{in}}'$ in each direction (N,S,E,W) by a different random scaling factor $f$, taken from a normal distribution with mean $1$ and standard deviation $\epsilon$. We look at the real and imaginary spectral gaps and coherence of the resulting model, averaged over different random configurations of transition rates. As shown in Fig. \ref{plt_SI_disorder}, the spectral gaps do not close for $\epsilon$ up to $80\%$, so the network is still in the topological regime. In general, the steady state is still localized on the edge of the state space, even though it can have a higher probability in a particular edge, e.g., the left edge, because of the non-uniform transition rates. The coherence of the oscillation is attenuated but some degree of oscillation persists. The robustness of the oscillations obtain from its topological nature, and we expect it to remain robust, up to some extent, for other types of perturbations such as additions of long-range interactions that connect nonadjacent states in the state space.

\section{The 2D Zak phase} \label{sec_Zak}

Our model topology is characterized by the topological invariant known as the 2D Zak phase \cite{liu2017novel,tang2021topology}. It is defined as an integral over the 2D Brillouin zone:
\begin{equation}
    \mathbf{\Phi}^c_i = \frac{1}{2\pi}\int_{BZ}\mathbf{A}_i(k_x,k_y) dk_x dk_y,
\end{equation}
where $\mathbf{A}_i(k_x,k_y)$ is the Berry connection for the $i$-th band, given by $\mathbf{A}_i(k_x,k_y)=i\langle \phi_i|\partial_\mathbf{k}|\psi_i\rangle$ for left and right eigenvectors $\langle\phi_i|$ and $|\psi_i\rangle$, respectively. For our stochastic system, we focus on the Zak phase $\mathbf{\Phi}^c_h$ for the highest band $h$ in real space (upper green band in Fig. \ref{plt_cohr_gap}b in the main text), which is the band closest to the steady state eigenvalue of 0. It turns out that when our system is in the trivial regime ($\rho<0$), we have $\mathbf{\Phi}^c_h=(0,0)$. When our system is in the topological regime ($\rho>0$), we have $\mathbf{\Phi}^c_h=(\pi,\pi)$. The emergence of the dynamical regime with edge currents and edge-localized steady state coincides with a nontrivial 2D Zak phase, indicating its topological origin. \newline

\section{Coherence}

In this section we motivate the definition of coherence in Eq. (\ref{eqn_cohr}) in the main text and discuss why it can serve as a measure for the robustness of oscillations. We also discuss how coherence changes with the parameters $\mu$ and $\rho$.

For a master equation with non-degenerate eigenvalues
\begin{equation}
    \frac{d\bp}{dt}=\cW\bp,
\end{equation}
the general solution is given by
\begin{equation}
    p_i(t) = \sum\limits_\nu c_\nu u_i^\nu e^{-\lambda_\nu t},
\end{equation}
where $\lambda_\nu$ are the eigenvalues, $(u_1^\nu, u_2^\nu,...,u_n^\nu)^T$ are eigenvectors corresponding to $\lambda_\nu$, and $c_\nu$ are constant coefficients that depend on initial conditions \cite{schnakenberg1976network}. The eigenvalues $\lambda_\nu$ of the transition matrix $\cW$ characterizes the timescales of the probability evolution for the corresponding eigenmodes. If the network represented by $\cW$ is irreducible and ergodic, then there exists a unique steady-state distribution $\bp^s$ such that $\cW\bp^s=0$, i.e., $\bp^s$ is an eigenvector of $\cW$ with a zero eigenvalue \cite{schnakenberg1976network}. The dynamics of $\bp(t)$, as we will see soon, is in general dominated by the first non-zero eigenvalue of $\cW$, which is the eigenvalue with the smallest modulus in the real part \cite{barato2017coherence}. Such eigenvalues generally come in conjugate pairs, denoted by $-\lambda_\text{R} \pm i\lambda_\text{I}$ for the real and imaginary parts $\lambda_\text{R},\lambda_\text{I}$, respectively. In particular, we take $\lambda_\text{R},\lambda_\text{I}\geq 0$ and $\lambda_1=-\lambda_\text{R} + i\lambda_\text{I}$.

\begin{figure*}[]
	\centering
	\includegraphics[width=6.6cm, height=5.2cm]{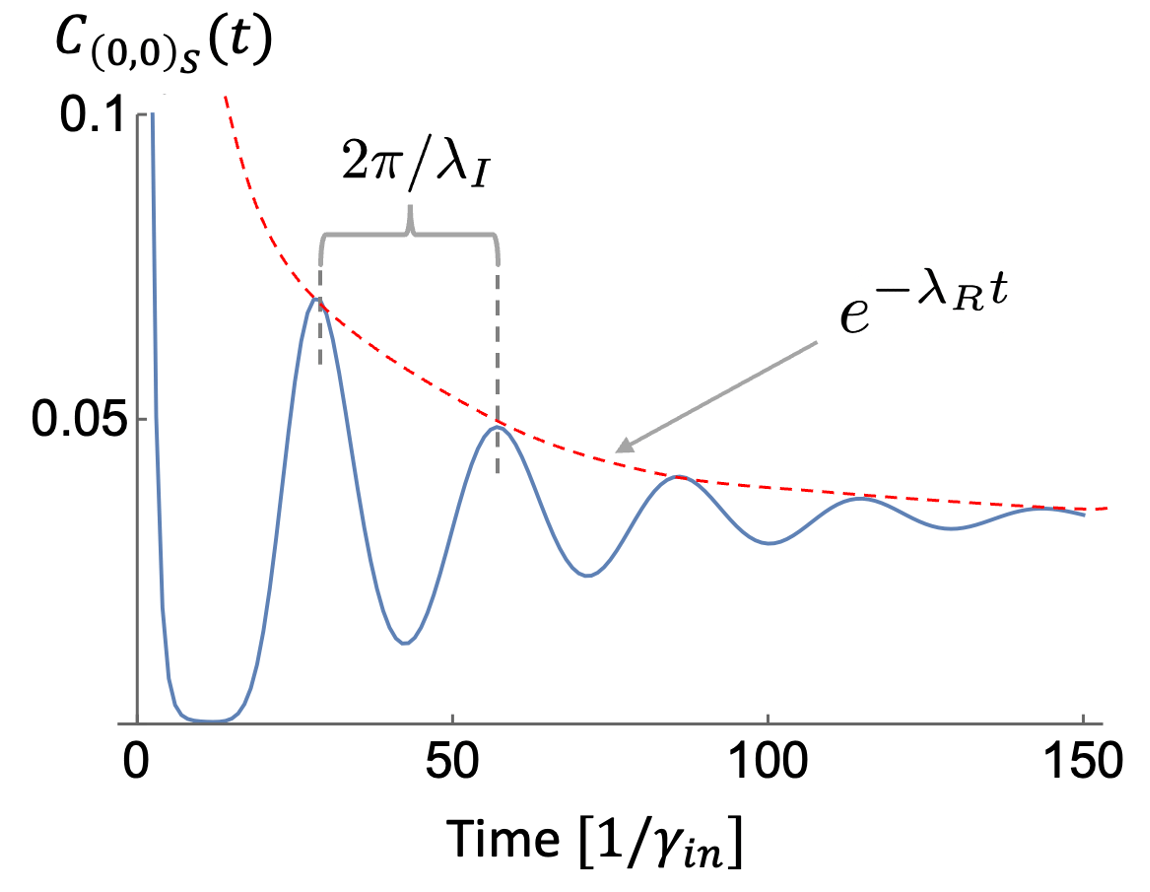}
	\caption{\textbf{Correlation function $\mathbf{C_{(0,0)_\text{S}}(t)}$ for the topological model goes through damped oscillations.} The oscillation has a decay time $\lambda_\text{R}^{-1}$ and period $2\pi/\lambda_\text{I}$. Coherence is defined as the ratio between these two timescales, characterizing the number of coherent oscillations that can be maintained before approaching steady-state. Eventually, $C_{(0,0)_\text{S}}(t)$ approaches the steady-state probability $p_i^s$. Time is plotted in units of $1/\gamma_{\text{in}}$. The parameters used are $\mu=7, \rho=5$.}
	\label{plt_SI_corrfunc}
\end{figure*}

The dynamics of the system can be studied through correlation functions $C_i(t)$. Following \cite{barato2017coherence}, we define $C_i(t)$ to be the probability to find the system at state $i$ at time $t$ given that the system starts at state $i$ at time $t=0$, i.e., $C_i(t) = p_i(t)$ given the initial condition $p_i(0)=1$ and $p_j(0)=0$ for all $j\neq i$. Starting from a state on the edge, say, the lower left corner $(0,0)_\text{S}$, the correlation function $C_{(0,0)_\text{S}}(t)$ goes through damped oscillations (see Fig. \ref{plt_SI_corrfunc}). After some transient behavior in the first oscillation cycle, the dynamics is dominated by $\lambda_1$, where the oscillation period is given by $T=2\pi/\lambda_\text{I}$ and the exponentially decaying envelope has a decay time $\lambda_\text{R}^{-1}$ \cite{barato2017coherence}. Therefore, we follow \cite{barato2017coherence} and define coherence as 
\begin{equation}
    \cR \equiv \frac{\lambda_\text{I}}{\lambda_\text{R}} ,
\end{equation} \label{eqn_SI_cohr}
which, when divided by $2\pi$, is the number of coherent oscillations that can be sustained before the system settles into the steady-state. The larger $\cR$ is, the more oscillations the system can maintain before stochastic fluctuations destroy its coherence. 

\begin{figure*}[]
	\centering
	\includegraphics[width=7cm, height=6.5cm]{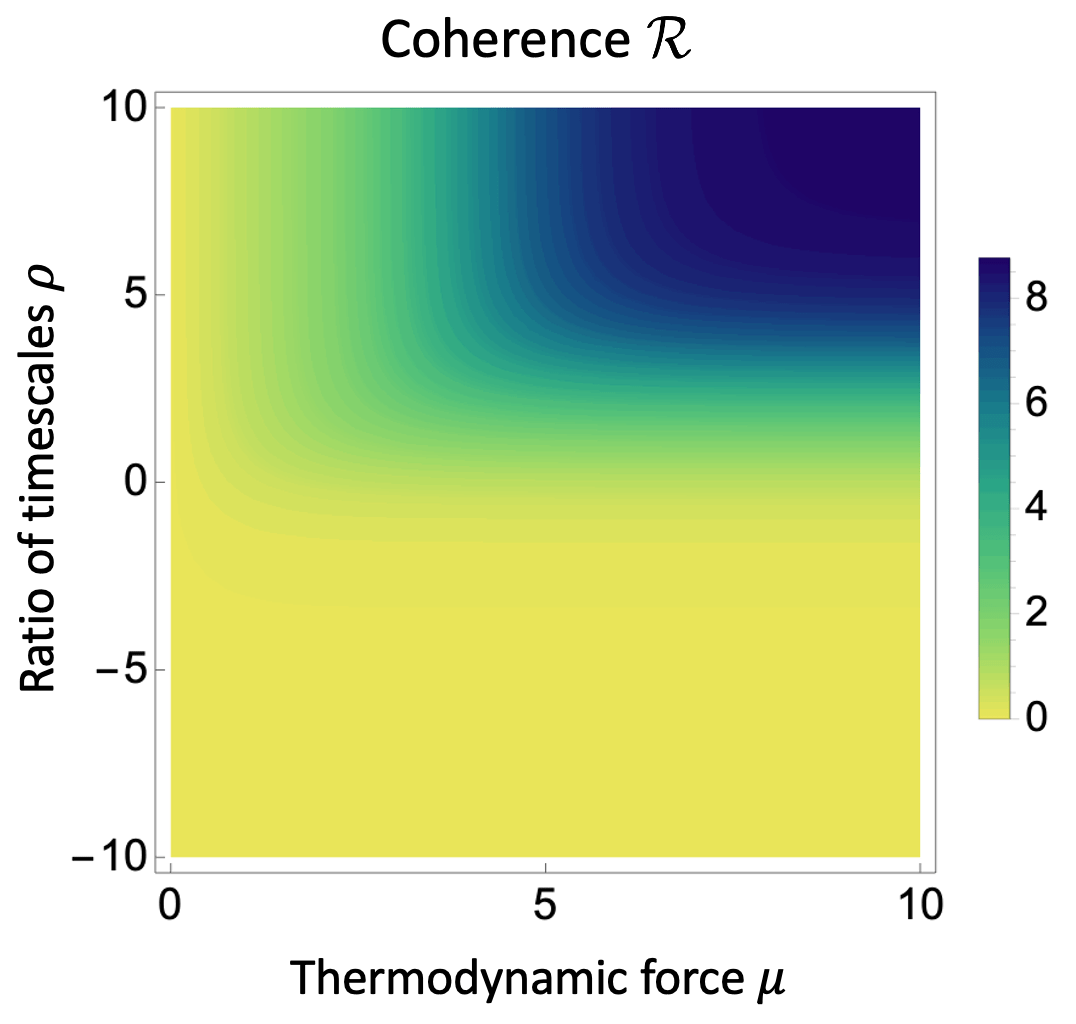}
	\caption{\textbf{Phase diagram for coherence.} The range of $\rho$ is extended from that of Fig. \ref{plt_coherence}a to include negative values. $\cR$ is monotonic in both $\mu$ and $\rho$, and remain close to 0 in the trivial regime $\rho<0$.}
	\label{plt_SI_cohr}
\end{figure*}

In Fig. \ref{plt_SI_cohr}, we show a phase diagram for $\cR$ that include negative values of $\rho$ (the trivial regime). $\cR$ turns out to be monotonic in both $\mu$ and $\rho$. $\cR$ increases monotonically in $\mu$ because a higher $\mu$ means a stronger external driving, making the forward reactions dominate more over their reverse reactions. This is more likely to give rise to trajectories with a particular chirality in our model. These trajectories correspond to the robust oscillations observed, and are less likely to backtrack or perform undirected diffusive motion when the external driving $\mu$ is large. For $\rho$, we can see that $\cR$ is always close to zero for $\rho<0$. Large values of coherence is obtained only with positive $\rho$, even if the thermodynamic force $\mu$ is large. This is expected because $\rho\sim 0$ is the transition that separates the topological regime from the trivial regime. In the trivial regime, the system does not support an edge state, and the dynamics resembles random diffusion in the bulk rather than a directed motion along the boundary. Since there are no global oscillations in this regime, the coherence remains low. When $\rho$ is positive and increasing, the system moves deeper into the topological regime, where the edge localization effects are more pronounced, leading to more robust oscillations.

\section{Single-molecule models for KaiC}

In this section, we include more details on the MWC model and the bilayer lattice model mentioned in the main text. Both models can be represented as directed networks and their dynamics described by corresponding master equations.

\begin{figure*}[]
	\centering
	\includegraphics[width=17cm, height=6cm]{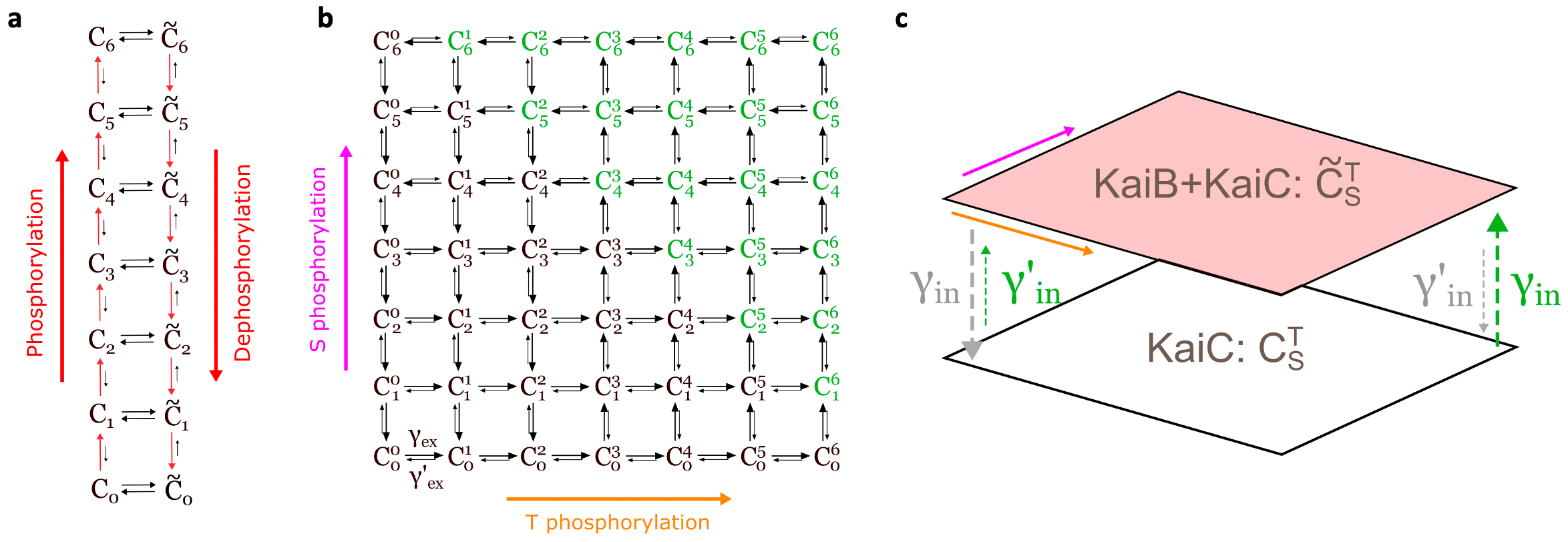}
	\caption{\textbf{Single-molecule models for KaiC.} \textbf{a}, The MWC model. The KaiC hexamer can be in one of two conformational states, the active state $C_i$ or the inactive state $\widetilde{C}_i$. KaiC tends to phosphorylate in the active state and dephosphorylate in the inactive state, highlighted by the vertical red arrows. The slower reverse reactions are represented by smaller black vertical arrows. Horizontal transitions correspond to conformational changes of the KaiC hexamer between active and inactive states. \textbf{b}, The bottom layer of the bilayer lattice model. Subscripts $x$ and superscripts $y$ represent T and S phosphorylation level, respectively, just as for the topological model. Solid black arrows represent phosphorylation/dephosphorylation transitions with rates $\gamma_{\text{ex}}$ and slower reverse rates $\gamma_{\text{ex}}'$ \textbf{c}, A zoomed-out view of the bilayer lattice model. Individual states in each layer are not shown. The green dashed arrows from the bottom to the top layer represent KaiB binding, while the gray dashed arrows from the top to the bottom layer represent KaiB unbinding. For the states colored green in \textbf{b}, transition rates are $\gamma_{\text{in}}$ for KaiB binding and $\gamma_{\text{in}}'$ for KaiB unbinding. For the states colored black, transition rates are $\gamma_{\text{in}}'$ for KaiB binding and $\gamma_{\text{in}}$ for KaiB unbinding. Larger dashed arrows (for any color) correspond to faster rates $\gamma_{\text{in}}$ while smaller dashed arrows correspond to slower rates $\gamma_{\text{in}}'$. During KaiB binding and unbinding, the phosphorylation levels remain the same.}
	\label{plt_SI_model}
\end{figure*}

The MWC paradigm \cite{monod1965nature} was first applied to the KaiABC system by \textcite{van2007allosteric}, which assumed all-or-none conformational change for the entire KaiC hexamer. By assumption, the KaiC hexamer can only be in two conformational states, the active states $C_i$ and the inactive states $\widetilde{C}_i$, where the subscript $0\leq i \leq 6$ represents the phosphorylation level of KaiC. There are 14 states in the state space in total. The network structure is shown in Fig. \ref{plt_SI_model}a. 

In this paper we follow one of the simplest versions of MWC-type models described in \cite{barato2017coherence} and use the same parametrization of the transition rates. When KaiC is in the active state, it is more likely to be phosphorylated. On the other hand, when KaiC is in the inactive state, it is more likely to be dephosphorylated. These dominant reactions are represented by the red vertical arrows in Fig. \ref{plt_SI_model}a with transition rates $\gamma e^{\eta/2}$. The reverse transitions (smaller black vertical arrows) have rates $\gamma e^{E/6}$. $\gamma$, $\eta$, and $E$ are model parameters that can be varied. For the horizontal transitions in Fig. \ref{plt_SI_model}a, the rates from $C_i$ to $\widetilde{C}_i$ are denoted as $k_{i\tilde{i}}$ and the rates from $\widetilde{C}_i$ to $C_i$ are denoted as $k_{\tilde{i}i}$, where
\begin{equation}
    k_{i\tilde{i}} = \begin{cases}
        1,  & i=0,1,2,3 \\
        e^{E(i-3)/3}, & i=4,5,6
    \end{cases}
\end{equation}
and 
\begin{equation}
    k_{\tilde{i}i} = \begin{cases}
        e^{E(3-i)/3},  & i=0,1,2,3 \\
        1, & i=4,5,6
    \end{cases}.
\end{equation}
The value of $\gamma$ sets the relative timescales between the phosphorylation/dephosphorylation reactions and conformational changes.

In Fig. \ref{plt_SI_model}b and \ref{plt_SI_model}c we show the network structure for the bilayer lattice model, adapted from \textcite{li2009circadian}. The state space consists of two layers of $7\times7$ lattices that are connected to each other. Fig. \ref{plt_SI_model}b shows the network structure of bottom layer from Fig. \ref{plt_SI_model}c. Similar to our topological model, the $x$ and $y$ coordinates (labeled by superscripts and subscripts $C^x_y$ in Fig. \ref{plt_SI_model}b) represent T and S phosphorylation levels, respectively. The top layer has an identical structure to the bottom layer, with states labeled by $\widetilde{C}^x_y$. $C^x_y$ in the bottom layer corresponds to unbound KaiC while $\widetilde{C}^x_y$ in the top layer represents KaiB-bound KaiC. The system can make transitions $C^x_y \ce{<=>} \widetilde{C}^x_y$ between layers while keeping $x$ and $y$ coordinates fixed, which corresponds to KaiB binding and unbinding. To aid comparison, we simplify the model by parametrizing the transition rates with $\mu$ and $\rho$, the same parameters for our topological model. Within both layers, the orientations for faster reactions on each edge are chosen such that they form an overall counterclockwise cycle, to capture the order of the phosphorylation cycle (see Fig. \ref{plt_SI_model}b). These phosphorylation/dephosphorylation reactions (solid black arrows) are assumed to have uniform rates $\gamma_{\text{ex}}$ while the slower reverse reactions have rates $\gamma_{\text{ex}}'$. On the other hand, the KaiB-binding transitions between layers have rates $\gamma_{\text{in}}$ and $\gamma_{\text{in}}'$. In the upper right half of the lattice where the states are labeled green in Fig. \ref{plt_SI_model}b, the dominant reaction between layers is the upward KaiB binding transitions (green dashed arrows in Fig. \ref{plt_SI_model}c) with rates $\gamma_{\text{in}}$, while the downward KaiB unbinding transitions (gray dashed arrows) have slower rates $\gamma_{\text{in}}'$. This is consistent with the fact that S phosphorylation promotes KaiB binding \cite{chow2022night}. In the rest of the lattice where the states are labeled black in Fig. \ref{plt_SI_model}b, the dominant reaction between layers is the downward KaiB unbinding transition with rates $\gamma_{\text{in}}$, while the upward KaiB binding transitions take the slower rate $\gamma_{\text{in}}'$. This parametrization scheme gives rise to cycles where a KaiC molecule phosphorylates in the bottom layer, binds to KaiB, gets dephosphorylated in the top layer, and unbinds with KaiB to restart the cycle. 

\section{Cost and precision for different KaiC models}

\begin{figure*}[]
	\centering
	\includegraphics[width=12.5cm, height=7.4cm]{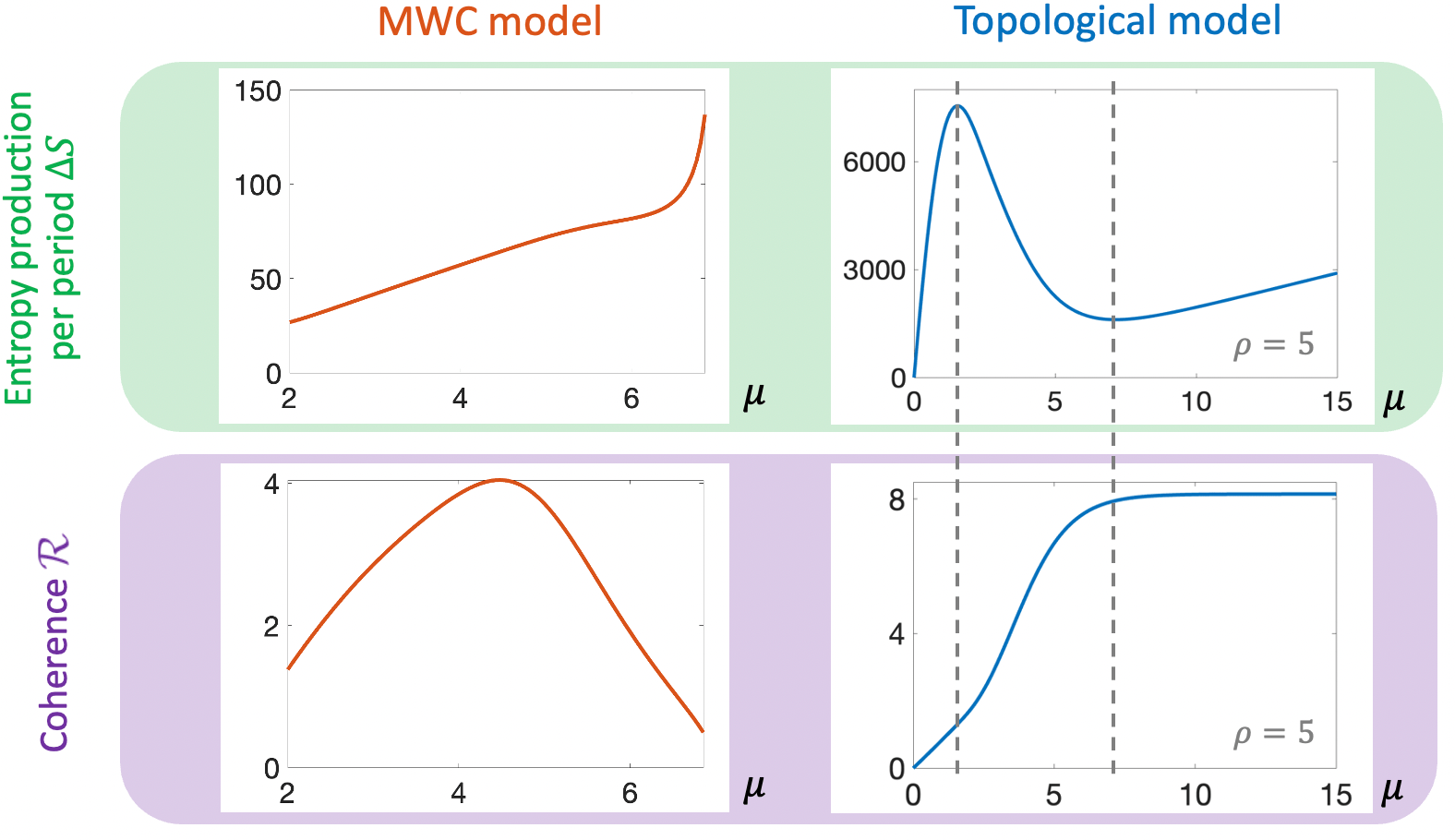}
	\caption{\textbf{Entropy production per period and coherence for the MWC and the topological model.} The MWC model requires more entropy production to maintain a higher coherence, and its coherence decreases as still more entropy is produced under a stronger driving $\mu$. On the other hand, the topological model supports a regime with increasing coherence but decreasing entropy production around $1.5<\mu<7$. For the MWC model, the parameters used are $\gamma=e^5$, $E=10$. We plot $\mu$ in a smaller range for this model because outside the plotted range, we have $\lambda_\text{I}=0$, which makes the oscillation period $\mathcal{T}$ and hence entropy production per period $\Delta S$ ill-defined. Entropy production per period is given in units of $k_B \gamma_{\text{tot}}$.}
	\label{plt_SI_cost_precision}
\end{figure*}

In this section we discuss how the free energy cost (quantified by $\Delta S$) and the precision (quantified by $\mathcal{R}$) of various KaiC models change with the external driving $\mu$. For the MWC model, the entropy production rate $\sigma$ of the network can be obtained by a cycle decomposition method \cite{schnakenberg1976network, barato2017coherence}:
\begin{equation}
    \sigma = \eta \gamma \sum\limits_{i=0}^5(e^{\eta/2}p^s_{C_i} - e^{E/6}p^s_{C_{i+1}}),
\end{equation}
where $p^s_{C_i}$ is the steady-state probability at the state $C_i$. This expression is used to calculate the entropy production per period $\Delta S = \sigma \mathcal{T}$ for the MWC model. As shown in the left column of Fig. \ref{plt_SI_cost_precision}, $\Delta S$ monotonically increases with $\cR$, as expected. Coherence, however, is not monotonic in $\mu$. When $\mu$ is sufficiently large, the MWC model ends up maintaining less coherent oscillations with increased driving and energetic cost. Note that the $\cR$ curve is different from the red dashed line shown in Fig. \ref{plt_coherence}e in the main text. Here we plot $\cR$ with $(\eta,E)$ held fixed, while the dashed line in Fig. \ref{plt_coherence}e shows the maximum coherence for each $\mu$ for any parameter combinations (also see Methods). In contrast to the MWC model, coherence for the topological model increases monotonically in $\mu$ for a fixed $\rho$. $\Delta S$, however, is non-monotonic, supporting a regime with increasing coherence and simultaneously decreasing cost (right column of Fig. \ref{plt_SI_cost_precision}). This unusual regime has its origin from the topological protection of the probability currents on the edge, which effectively reduces the state space into a one-dimensional cycle along the edge. This edge localization leads to a lower free energy cost but more coherent oscillations. 
\section{Spectral gap as a predictor of coherence} \label{sec_SI_gap}

In this section we study the effects of $\mu$ and $\rho$ on the spectral gaps of the transition matrix $\cW$ with both periodic boundary conditions (PBC) and open boundary conditions (OBC), by tracking how the eigenvalues change with the two parameters. We also discuss why the spectral gap can serve as a predictor of coherence.

\begin{figure*}[]
	\centering
	\includegraphics[width=17cm, height=7.1cm]{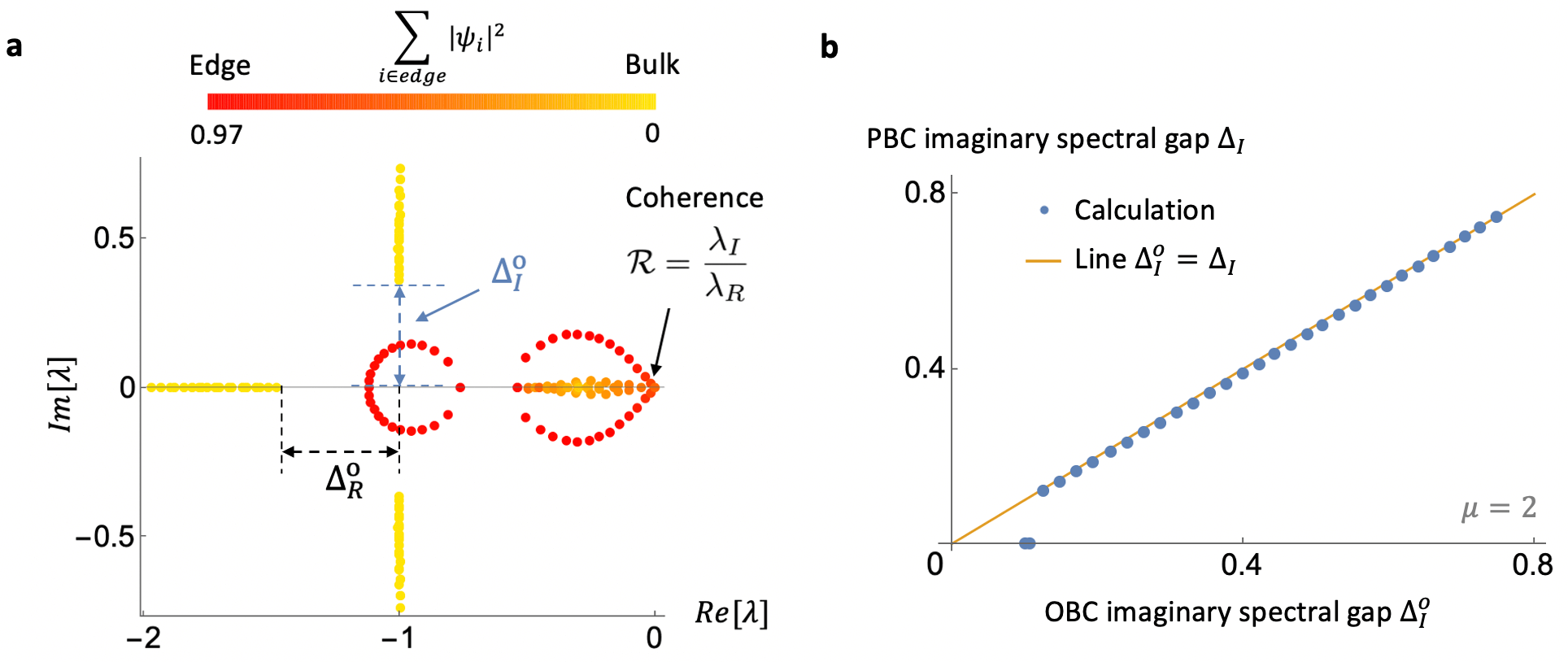}
	\caption{\textbf{Spectral properties of $\cW$ in open boundary conditions (OBC).} \textbf{a}, Spectrum of $\mathbf{\cW}$ in OBC. The real spectral gap is defined as the shortest distance between the left band to the imaginary axis, and the imaginary spectral gap is defined as the shortest distance between the top band to the real axis. The color for each dot corresponds to the value of $\sum\limits_{i\in\text{edge}}|\psi_i|^2$, the sum of the magnitude squared of all entries in the steady-state eigenvector that lie on the edge. This quantity measures the extent to which the corresponding eigenvector is localized on the boundary of the system, and lies on a continuous spectrum between 0 and 1. A redder color means more localization. The solid black arrow points at the eigenvalue $\lambda_1=-\lambda_\text{R}+i\lambda_\text{I}$, which determines the coherence $\cR$. The parameters used are $\mu=2, \rho=1$. \textbf{b}, The imaginary spectral gaps defined in PBC and OBC are nearly identical when both gaps are large and the bands are well separated. $\rho$ is varied from $0.2$ to $4.6$ in the plot while $\mu$ is fixed at $\mu=2$. Calculated values for the two gaps ($\Delta_\text{I}$ in PBC and $\Delta_\text{I}^\text{o}$ in OBC blue dots) almost all lie on the orange line on which the two gaps are equal. Eigenvalues and spectral gaps are given in units of $\gamma_{\text{tot}}$.}
	\label{plt_SI_OBC}
\end{figure*}

Spectral gaps can be defined for both PBC and OBC. Taking PBC, we can write the transition matrix in reciprocal space as 
\begin{equation} \label{eqn_W}
    \cW_{\mathbf{k}} =
\begin{pmatrix}
    -\gamma_{\text{tot}} & \gamma_{\text{in}} + \gamma_{\text{ex}}' e^{-i k_y} & 0 & \gamma_{\text{in}}' + \gamma_{\text{ex}} e^{-i k_x}\\
    \gamma_{\text{in}}' + \gamma_{\text{ex}} e^{i k_y} & -\gamma_{\text{tot}} & \gamma_{\text{in}} + \gamma_{\text{ex}}' e^{-i k_x} & 0\\
    0 & \gamma_{\text{in}}' + \gamma_{\text{ex}} e^{i k_x} & -\gamma_{\text{tot}} & \gamma_{\text{in}} + \gamma_{\text{ex}}' e^{i k_y} \\
    \gamma_{\text{in}} + \gamma_{\text{ex}}' e^{i k_x} & 0 & \gamma_{\text{in}}' + \gamma_{\text{ex}} e^{-i k_y} & -\gamma_{\text{tot}}
\end{pmatrix}.
\end{equation}
We obtain the spectral gaps for PBC from the continuous spectrum defined by $\cW_{\mathbf{k}}$, as illustrated in Figs. \ref{plt_cohr_gap}a and \ref{plt_cohr_gap}b in the main text. Similarly, we can obtain spectral gaps for OBC by plotting the spectrum of $\cW$ in the complex plane, as in Fig. \ref{plt_SI_OBC}a. For each eigenstate $\psi$, we also calculate $\sum\limits_{i\in\text{edge}}|\psi_i|^2$, the sum over the squared magnitudes $|\psi_i|^2$ for all entries $i$ that lie on the edge of the system. This quantity characterizes how localized the corresponding eigenstate is on the edge. We normalize each eigenstate so that the value of $\sum\limits_{i\in\text{edge}}|\psi_i|^2$ can vary continuously between 0 and 1. A larger sum, or a redder color, indicates more edge localization. As we can see from Fig. \ref{plt_SI_OBC}a, the OBC spectrum displays two circle-like shapes with eigenstates highly localized on the edge and four clusters of eigenvalues, lying on the real or imaginary axis, with eigenstates more localized in the bulk. These four clusters are identified as the ``bands'' in the OBC case, in analogy with the four bands in the PBC spectrum as shown in Fig. \ref{plt_cohr_gap}a and \ref{plt_cohr_gap}b in the main text. As shown in Fig. \ref{plt_SI_OBC}a, we define the OBC real spectral gap $\Delta_\text{R}^\text{o}$ as the shortest distance from the leftmost band to the imaginary axis and the OBC imaginary spectral gap $\Delta_\text{I}^\text{o}$ as the shortest distance from the topmost band to the real axis.

\begin{figure*}[]
	\centering
	\includegraphics[width=17cm, height=11.5cm]{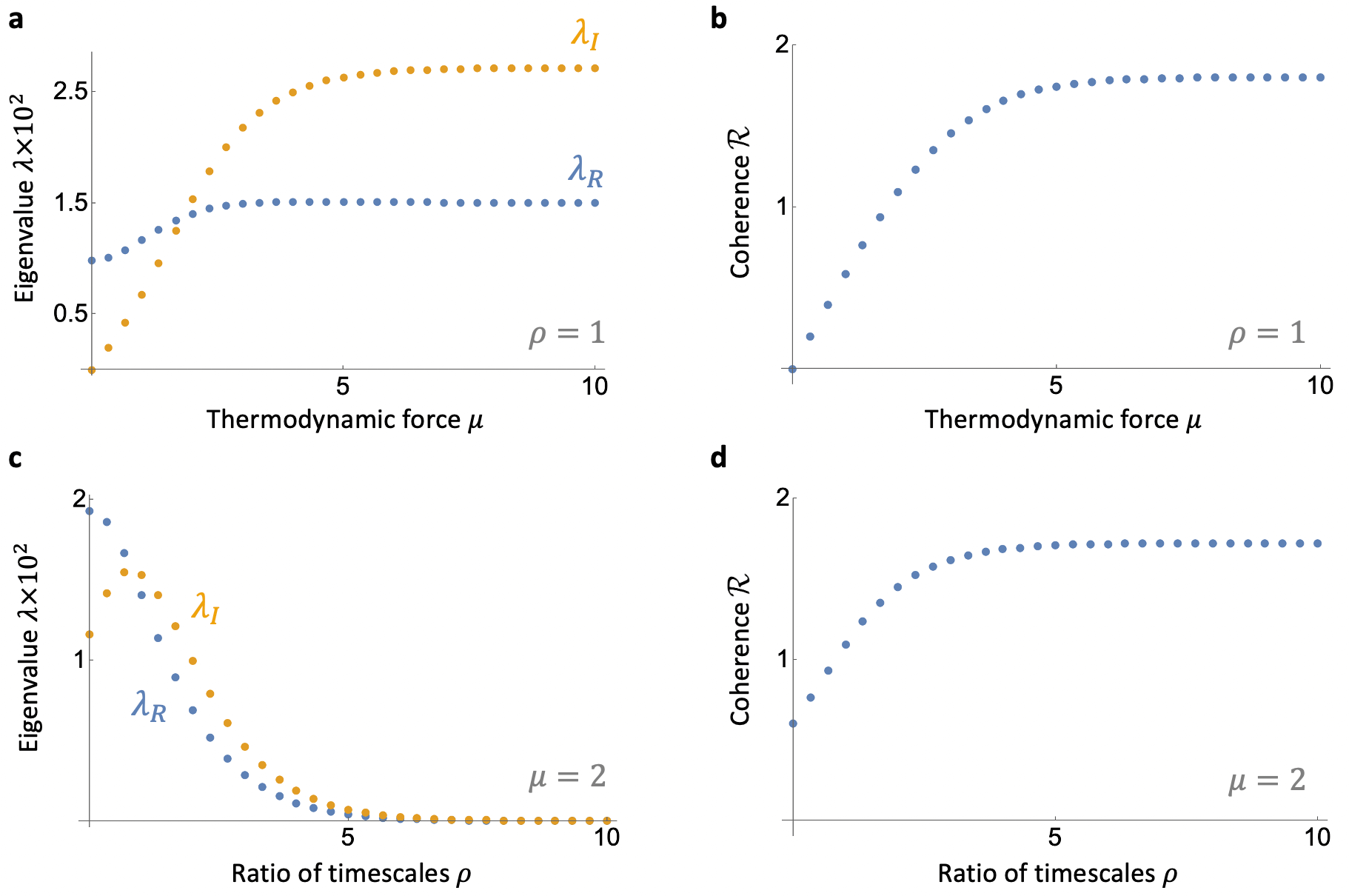}
	\caption{\textbf{Effects of $\mathbf{\mu}$ and $\mathbf{\rho}$ on the first non-zero eigenvalue $\mathbf{\lambda_1}$ and coherence.} \textbf{a}, The absolute value of the real and imaginary parts for $\lambda_1$ as a function of $\mu$. $\lambda_\text{I}$ increases with $\mu$ and approaches a maximum while $\lambda_\text{R}$ does not change significantly. \textbf{b}, Coherence as a function of $\mu$. $\cR$ is almost proportional to $\lambda_\text{I}$ given a nearly constant $\lambda_\text{R}$. \textbf{c}, The absolute value of the real and imaginary parts of $\lambda_1$ as a function of $\rho$. $\lambda_\text{R}$ decreases monotonically while $\lambda_\text{I}$ is non-monotonic in $\rho$. In the regime where $\lambda_\text{I}$ decreases, it decreases slower than $\lambda_\text{R}$. \textbf{d}, Coherence as a function of $\rho$, showing a monotonic increase just as in the $\mu$ direction. Eigenvalues are given in units of $\gamma_{\text{tot}}$.}
	\label{plt_SI_lambda_cohr}
\end{figure*}

In Fig. \ref{plt_SI_OBC}b we compare the imaginary spectral gap for PBC and OBC for the same parameters $\mu$ and $\rho$. The two gaps are almost identical at large $\rho$ when the gap is large. When $\rho$ is closer to 0, overlaps between the bands begin to develop for both PBC and OBC, which renders the spectral gap ill-defined. In the regime of relatively large $\rho$, spectral gaps for the OBC and PBC can be used interchangeably. The same conclusion can be drawn by looking at the two gaps with changing $\mu$ and fixed $\rho$. Therefore, to understand the relationship between coherence and the spectral gap, we can look at the OBC spectrum instead.

Studying how $\lambda_1$ and the entire OBC spectrum change with $\mu$ and $\rho$ gives a better picture of why coherence and the imaginary spectral gap track each other. Varying $\mu$ and $\rho$ leads to global changes in the spectrum of $\cW$ in the complex plane, such as narrowing of the bandwidth or movement of all eigenvalues in a band in the same direction. Meanwhile $\lambda_1$, the particular eigenvalue that determines coherence (see Fig. \ref{plt_SI_OBC}a), moves along with the bands in roughly the same way. When $\rho$ is fixed, increasing $\mu$ has the effect of increasing the range or dispersion of the spectrum in the imaginary part. When $\mu=0$, all reactions have symmetric rates with respect to their reverse reactions, and the system is in detailed balance. In this case, the spectrum is entirely real because $\cW$ is symmetric. Increasing $\mu$ therefore introduces nonzero imaginary parts to the eigenvalues, leading to oscillatory modes with faster timescales as the driving increases. For the top band, for instance, the imaginary parts of all eigenvalues in the band move upwards in the complex plane by roughly the same distance, keeping the bandwidth constant but increasing the distance between the bottom point in the band to the real axis (which is the definition of $\Delta_\text{I}^\text{o}$). Meanwhile, the real parts of the eigenvalues in the four bands remain virtually unchanged with $\mu$. As shown in Fig. \ref{plt_SI_lambda_cohr}a, $\lambda_1$ moves in a similar way with the global spectrum. Increasing $\mu$ leads to an increase in $\lambda_\text{I}$ but very little change in $\lambda_\text{R}$. Since $\cR$ is proportional to $\lambda_\text{I}$ while $\lambda_\text{R}$ is roughly constant, coherence follows the same functional relationship as $\lambda_\text{I}$ when $\mu$ increases (Fig. \ref{plt_SI_lambda_cohr}b). Therefore, since $\Delta_\text{I}$ closely tracks $\lambda_\text{I}$ (see Fig. \ref{plt_cohr_gap}c in the main text), it also closely tracks $\cR$ monotonically.

On the other hand, $\rho$ serves as an overall compression factor, decreasing the bandwidth for all four bands. The points farthest away from the real axis for the imaginary bands also increase slightly with $\rho$. The absolute values of these points equal the sum of the bandwidth for the imaginary band and the imaginary spectral gap. Therefore, a decrease in the bandwidth and an increase in the sum of the bandwidth and the spectral gap imply a monotonic increase in the spectral gap with increasing $\rho$. $\lambda_1$ in this case, however, does not always follow the general movement of the global spectrum. As shown in Fig. \ref{plt_SI_lambda_cohr}c, $\lambda_\text{R}$ still decreases monotonically as expected, as the rightmost band is compressed toward the origin. $\lambda_\text{I}$, on the other hand, develops non-monotonic behavior. Nevertheless, when $\lambda_\text{I}$ begins to decrease, it approaches zero more slowly than $\lambda_\text{R}$ and gives rise to a monotonically increasing $\cR$ that tracks $\Delta_\text{I}^\text{o}$ (and therefore $\Delta_\text{I}$) in $\rho$ (Fig. \ref{plt_SI_lambda_cohr}d). 

\section{Generalizations to many molecules} \label{sec_many_molecules}
In this section, we explore generalizations of our single-molecule model to many molecules. In particular, we investigate the higher-dimensional state space formed by two or more KaiC molecules. We further assume that KaiA is a scarce resource and study the resulting state space geometry modified by the constraints introduced by the competition over KaiA. This competition is intrinsically a population-level effect, as all KaiC molecules compete for the same scarce resource to complete the phosphorylation cycle. Finally, we look at the effects of limited or excess amounts of KaiA on the phosphorylation cycles for our single-molecule model or the many-molecule generalization.

We first look at the state space of our model when there are two KaiC molecules in consideration. In this case, the indices $(x_1,x_2,y_1,y_2)_{s_1,s_2}$ completely specify the system of two molecules, where the subscripts on the variables $x,y,s$ denote molecule number. For each phosphorylation level $(x_1,x_2,y_1,y_2)$, there are $4\times4=16$ internal states that form a tesseract, the 4D generalization of a cube. These internal states repeat themselves in the four directions spanned by the external variables $x_1,x_2,y_1,y_2$ to form a state space in the shape of a 4D hypercube.

\begin{figure*}[]
	\centering
	\includegraphics[width=16cm, height=10.7cm]{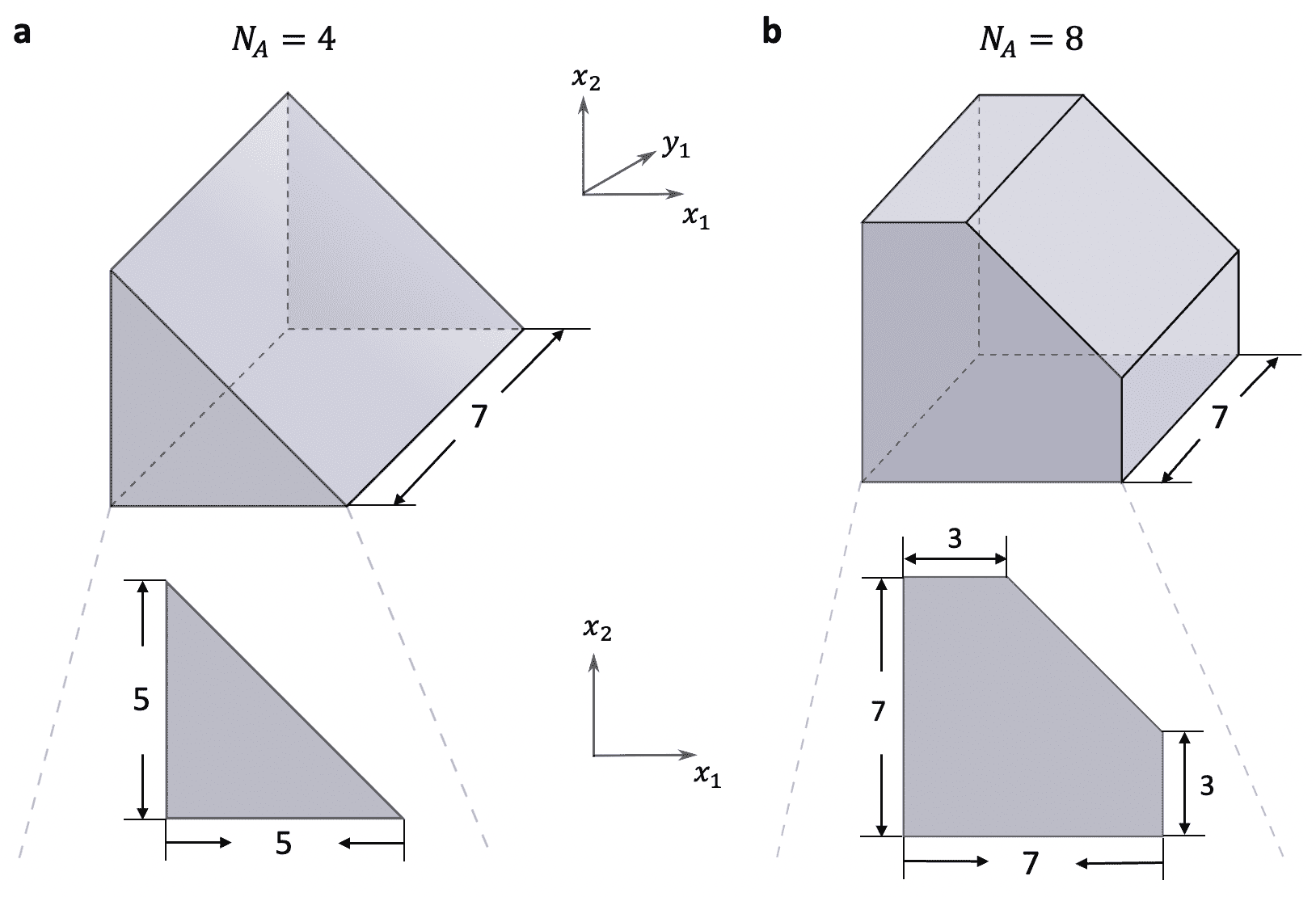}
	\caption{\textbf{Cross sections of the 4D state space for two KaiC molecules when competition for KaiA is taken into account.} \textbf{a}, For $N_\text{A}=4$, the constraint $x_1+x_2\leq N_\text{A}$ from competition over KaiA reduces the state space in $x_1$-$x_2$ from a $7\times 7$ square to an isosceles right triangle with side length $5$. All side lengths refer to the number of phosphorylation levels contained in that direction. The 3D cross section in $y_2$ for the complete 4D state space is a right triangular prism, obtained by repeating the isosceles right triangle in $y_1$. Such geometries come up in the regime $N_\text{A}\leq 6$. \textbf{b}, For $N_\text{A}=8$, the constraint reduces the state space to an irregular pentagon in $x_1$-$x_2$, which results from cutting off a corner from a $7\times 7$ square. The 3D cross section in $y_2$ is a pentagonal prism that derives from repeating the pentagon in $y_1$. Such geometries come up in the regime $6<N_\text{A}<12$.}
	\label{plt_SI_state_space}
\end{figure*}

To study population-level effects, we assume that KaiA molecules, which promotes KaiC phosphorylation \cite{xu2003cyanobacterial}, are a scarce resource that can be depleted. This agrees with experimental observations that when KaiA concentrations drop below a certain point, oscillations are prohibited and KaiC phosphorylation levels remain low \cite{kageyama2006cyanobacterial}. We further assume that each KaiA molecule promotes the T phosphorylation of one KaiC monomer. Since there are six T phosphorylation sites for each KaiC hexamer, we consider the regime $N_\text{A}<12$, where the number of KaiA molecules $N_\text{A}$ is not enough to fully phosphorylate both molecules. This competition introduces the constraint $x_1+x_2\leq N_\text{A}$. In the following, we consider the geometry of this state space in two cases. In either case, the hypercubic state space of the system is reduced to a hyperprism from the constraint introduced by competition over KaiA. Depending on the amount of available KaiA, the cross-sections of the hyperprism can be different, as illustrated in Fig. \ref{plt_SI_state_space}. Such changes to the geometry of the state space modifies its edges and introduces dependence on phosphorylation levels of different molecules along these new edges, e.g., $x_1+x_2=4$ on the hypotenuse of the triangle in Fig. \ref{plt_SI_state_space}a.

One regime is when $N_\text{A}\leq 6$. This is when KaiA can fully phosphorylate one KaiC hexamer at most. In this case, the $x_1$-$x_2$ subspace will be reduced from a 2D square to an isosceles right triangle, as illustrated in the lower part of Fig. \ref{plt_SI_state_space}a, taking $N_\text{A}=4$ as an example. To study the shape of the four-dimensional state space, we can look at its 3D cross-sections by holding one coordinate fixed. As shown in Fig. \ref{plt_SI_state_space}a, with a fixed $y_2$, the state space in $x_1$-$x_2$-$y_1$ space is a right triangular prism, obtained by repeating the right triangle in $x_1$-$x_2$ along the $y_1$ direction. This 3D cross-section is the same for any $y_2$, which means that this prism repeats itself in the fourth dimension $y_2$, forming a 4D hyperprism.

In another regime, $6<N_\text{A}<12$. This is when KaiA can fully phosphorylate one molecule but not enough to phosphorylate both. In Fig. \ref{plt_SI_state_space}b, we illustrate the geometry for the case $N_\text{A}=8$. The $x_1$-$x_2$ subspace is an irregular pentagon, formed by removing the upper right corner (an isosceles right triangle) from the 2D square. The 3D cross-section for a fixed $y_2$ is a pentagonal prism that repeats in $y_2$ to form another 4D hyperprism, which differs from the above by its different 3D cross-section.

In general, if there are $N$ KaiC molecules, the constraint on the T phosphorylation levels becomes $x_1+x_2+...+x_N\leq N_\text{A}$. The state space will be a $2N$-dimensional hyperprism, where high-dimensional ``corners'' are removed from a $2N$-dimensional hypercube due to the constraint.

\begin{figure*}[]
	\centering
	\includegraphics[width=14cm, height=10cm]{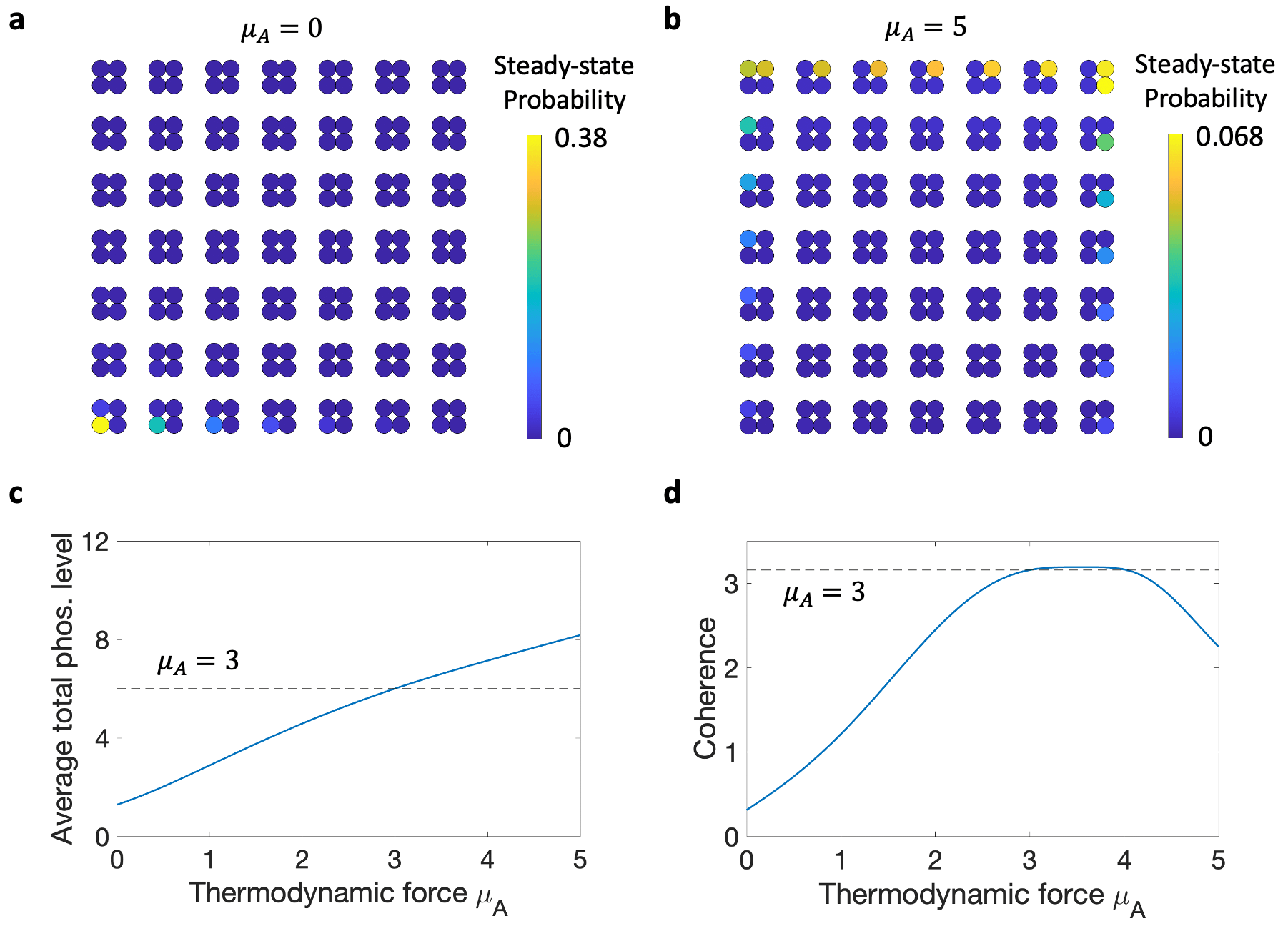}
	\caption{\textbf{The effects of limited or excess amounts of KaiA on the KaiC phosphorylation cycle.} \textbf{a}, Steady-state probability distribution in the absence of KaiA. The thermodynamic force for the $S\rightarrow E$ internal transition, $\mu_\text{A}$, which characterizes the external driving from interaction with KaiA, is set to be 0. The probability is localized on the lower left corner, corresponding to a hypophosphorylated state for KaiC in the absence of KaiA, consistent with experimental results \cite{kageyama2006cyanobacterial}. In all panels, we take $\mu=3,\rho=5$. \textbf{b}, Steady-state probability distribution for $\mu_\text{A}=5$. The probability is localized in highly phosphorylated states of KaiC, also consistent with experiments \cite{chavan2021reconstitution}. \textbf{c}, Total phosphorylation level of T and S as a function of $\mu_\text{A}$, averaged over the steady-state probability distribution. KaiC autodephosphorylates as $\mu_\text{A}$ decreases with the removal of KaiA, and phosphorylates as $\mu_\text{A}$ increases with the addition of KaiA. $\mu_\text{A}=\mu=3$ corresponds to the unmodified model as introduced in the main text, whose phosphorylation level is illustrated with a horizontal dashed line. \textbf{d}, Coherence as a function of $\mu_\text{A}$. Coherent oscillations are inhibited with the removal of KaiA or an excess of KaiA. Coherence for $\mu_\text{A}=3$ is illustrated with a horizontal dashed line.}
	\label{plt_SI_noA}
\end{figure*}

When the amount of KaiA is varied, this simple model behaves the same as what is observed in experiments, i.e., limited KaiA leads to sustained low levels of phosphorylation while excess KaiA leads to sustained high levels of phosphorylation \cite{kageyama2006cyanobacterial,chavan2021reconstitution}. When there is little to no KaiA, states with high T phosphorylation levels are blocked. When there is too much KaiA, there is no competition and no constraint whatsoever on the state space, which means that each KaiC molecule oscillates independently without synchronizing. Moreover, the strong driving from KaiA would promote phosphorylation of each molecule (see Fig. \ref{plt_SI_noA}b), leading to a high overall phosphorylation level. We note that the above discussions have relied on simplifying assumptions on the nature of KaiA-KaiC interactions.  A more thorough investigation of a many-molecule model would take into account more realistic properties of KaiA.

Such effects of KaiA can also be captured just by our single-molecule model in the main text. Because the thermodynamic force for the $S\rightarrow E$ internal transition, which we denote as $\mu_\text{A}$, characterizes the driving from interaction with KaiA, changes in $\mu_\text{A}$ correspond to changes in concentrations of KaiA \cite{hill1989}. We keep the slower $E\rightarrow S$ transition rate fixed at $\gamma_{\text{in}}'$ and let the $S\rightarrow E$ transition rate depend on $\mu_\text{A}$ by $\gamma_{\text{in}}'e^{\mu_\text{A}/k_B T}$. Fig. \ref{plt_SI_noA}a shows the steady-state probability distribution for $\mu_\text{A}=0$ while $\mu$ and $\rho$ are kept the same, which corresponds to complete removal of KaiA. As we can see, the probability is mostly localized in states near the lower left corner, corresponding to low levels of phosphorylation. In contrast, in Fig. \ref{plt_SI_noA}b we set $\mu_\text{A}=5$, corresponding to excess amounts of KaiA. The probability, in turn, becomes localized in highly phosphorylated states. The average total phosphorylation level (T and S combined) as a function of $\mu_\text{A}$ is shown in Fig. \ref{plt_SI_noA}c, which shows increasing levels of phosphorylation as $\mu_\text{A}$ increases and KaiA is added. For either limited ($\mu_\text{A}<3$) or excess ($\mu_\text{A}>3$) amounts of KaiA, oscillation tends to be attenuated as illustrated by the decrease in coherence in Fig. \ref{plt_SI_noA}d.

\bibliography{references}